\documentclass{llncs}

\usepackage[cmex10]{amsmath}
\usepackage{amssymb}
\usepackage{latexsym}

\usepackage{graphicx}
\usepackage{color}
\usepackage{stfloats}

\usepackage{ntheorem}
\usepackage{graphicx}
\usepackage{paralist}
\usepackage{pstricks}
\usepackage{array}
\usepackage{comment}

\newlength{\mylengtha}
\newlength{\mylengthb}

\newcommand{\path}{\omega}

\newcommand{\probkn}{P_{\neg {\it drop}}}

\newcommand{\probsn}{P_{{\it select}}}
\newcommand{\probs}{\frac{s}{c}}

\newcommand{\probnotsn}{P_{\neg {\it select}}}
\newcommand{\probdn}{P_{{\it drop}}}

\newcommand{\proboch}{P_{inx}}

\newcommand{\sminus}{\backslash}
\newcommand{\sasb}{\widehat{s}}
\newcommand{\chat}{\widehat{c}}

\newcommand{\probls}{P_{\it loss}}
\newcommand{\probnls}{P_{\neg {\it loss}}}

\newcommand{\items}{n}
\newcommand{\buffer}{s}
\newcommand{\cache}{c}
\newcommand{\theitem}{d}

\newcommand{\gp}{gossip}

\newcommand{\msg}{\sigma}

\newcommand{\ie}{i.e.}

\newcommand{\revision}[1]{{#1}}

{\theoremstyle{definition}
\theorembodyfont{\rmfamily}
\newtheorem*{scenario}{State}}
\newcommand{\oper}[1]{{\sf\scriptsize #1}}

\title{
 A Modeling Framework for Gossip-based Information Spread
}

\author{
  Rena Bakhshi\inst{1}
\and
  Daniela Gavidia\inst{2}
\and
  Wan Fokkink\inst{1}
\and Maarten van Steen\inst{1}
}

\institute{
  Vrije Universiteit Amsterdam,
  Department of Computer Science,\\
  De Boelelaan 1081a,
  1081 HV Amsterdam,
  The Netherlands\\
  {\textbraceleft {\tt rbakhshi,wanf,steen}\textbraceright{\tt@few.vu.nl}}
\and 
 Chess, 
 P.O. Box 5021, 2000 CA Haarlem \\
  {\tt dgavidia@gmail.com}
}
\begin{document}

\maketitle
\pagestyle{plain}
\pagenumbering{arabic}

  \begin{abstract}
  We present an analytical framework for gossip protocols based on the pairwise 
  information exchange between interacting nodes. This framework allows 
  for studying the impact of protocol parameters on the performance of 
  the protocol. Previously, gossip-based information dissemination protocols have been analyzed under the assumption 
  of perfect, lossless communication channels. We extend our framework for the analysis of   
  networks with lossy channels.
 We show how the presence of message loss, coupled with specific topology configurations, impacts the expected 
 behavior of the protocol. 
 We validate the obtained models against simulations for two protocols. 
  \end{abstract}

\section{Introduction}

Gossip protocols have emerged as an attractive solution for distributing 
information in large-scale systems, due to their simplicity and efficiency. 
Randomness and the distributed nature of gossip protocols considerably 
increase their robustness compared to deterministic protocols, also in the presence of 
failures and data loss. 

Although gossip protocols are characterized by their superficial simplicity,  
the large-scale behavior of a gossip system is not easily predictable. 
From the perspective of protocol design and behavior prediction, this is an 
undesirable situation, and with the growing number of gossip protocols, 
there is an increasing demand for their analysis in an insightful and systematic way. 

However, the formal analysis of these protocols is still a rather unexplored 
research field with many challenges, in part caused by the fact that traditional 
techniques quickly lead to a state-space explosion (see, e.g., \cite{FG06,1481510,bakhshi:phd}). 
In \cite{BGFSCN09}, we have proposed a modelling framework for gossip-based 
information dissemination protocol, in which each node periodically 
selects a random peer and shuffles its local data. In this paper, we show that 
this approach is generally applicable to gossip protocols for information 
dissemination, and we extend the analysis method to networks with lossy channels. 
An exchange of data between nodes is modelled as a state transition capturing 
the presence or absence of a selected data item before and after a gossip between 
two nodes. 

Our modelling framework allows for the prediction of the system behavior in 
large-scale environments. We propose a model based on local pairwise interactions
which in combination with model checking (cf.~\cite{BF09}) or with the mean-field framework (cf.~\cite{BCFH:10})
allows for the analysis of the emergent behaviour of the system.
The model can also be used for optimization of system parameters and performance (cf.~\cite{BGFSCN09}), 
and for fast event-based simulation, as it is done in this paper.

We introduce our theoretical framework initially 
assuming no communication failures in the network. Later, we  
show that the framework is applicable to networks where nodes communicate 
over an unreliable medium. 
An important part of our proposed framework is the probability $\probdn$ that 
an observed item in a node's local storage has been replaced by another item after 
a gossip. We derive the expressions for this probability under 
the assumption of perfect communication medium for two protocols, Newscast~\cite{newscast:tr} and Shuffle~\cite{GVS06}. 
In the presence of message loss, however, the dependence of the probability on 
the type of underlying network graph emerges. We deconstruct the expression for $\probdn$ 
into its main components, and identify the ones that depend on specific scenario 
configurations, \ie, message loss and topology combinations. One such component is 
the probability of an observed item found at one gossiping node, to be also found at its peer. 
Since it is not feasible to cover all 
possible network topologies in one formula, the computation of this probability 
is based on statistical data that can be obtained from Monte Carlo 
simulations of the protocol at hand.

Since the work of Demers et al.~\cite{41841}, \gp{} protocols often 
follow bidirectional data exchange (push-pull) for better performance. As observed in 
\cite{DONB07}, the performance of these protocols is usually evaluated under the assumption of a perfect, 
error-free communication medium. Thus, data exchange operations between nodes are generally  
considered to be atomic operations, e.g.,~\cite{649816,1148679,SRS02,JKS03,VoulgarisGS05,GVS06,1127791,BGFSCN09}. 
That is, if a node initiates a gossip with another node, both nodes base their 
local decisions upon each other's data, and in some protocols 
even guarantee the preservation of each other's data. In practice, however, implementing data exchange as 
an atomic operation is hard to achieve assuming communication over unreliable media~\cite{559404}.

In this respect, our paper makes an additional contribution; it investigates the impact of a realistic 
environment with lossy communication channels on a push-pull-based \gp{} protocol.  
 Furthermore, we present in more detail the following observations: 
\begin{compactitem}[$\bullet$]
 \item Introducing message loss in the network model affects the emergent behavior 
 of the protocol in a specific manner: with the introduction of lossy communication channels, 
 the correlation between local storage content of neighboring nodes increases, and the degree of the correlation 
 depends on the network topology. 
 \item The fewer neighbors nodes have in the underlying network, the stronger 
 the effect of message loss on the emergent behavior of the protocol.
 For fully connected networks, message loss does not have an impact on the 
 distribution of data over the network, which remains uniform (as observed when there are no losses).
\end{compactitem}

%

Two areas of research are relevant to our work: generic modelling frameworks for 
gossip protocols and the performance of gossip protocols in the presence of 
message loss. 
Automated mean-field framework for dynamic gossip networks has been presented 
in \cite{BEEFH10}. Due to the underlying mean-field method, used 
by this framework, the results are accurate only for very large networks and for average 
behavior of the modelled system. The mean-field framework can be combined with 
our framework, as it is shown in \cite{BCFH:10} for Shuffle.

There is previous work \cite{1582743} on a simple gossip-based membership protocol with 
nonatomic protocol actions in the presence of message loss of up to 1\%. 
The authors proposed to use a push-based gossip protocol, in which only an active node sends data to its peer, 
and immediately removes the sent data from its cache. Moreover, the node sends only two 
items, IP addresses and ports of itself and of a random peer from its local cache. In the protocols that we study, 
each gossiping pair exchanges a random subset of the data items. The authors show the protocol correctness 
modelling it by graph transformations, similar to the approach in \cite{1481510}. The protocol properties 
analyzed include the expected number of neighbors a node has and the uniformity of the neighborhood lists. 
The theoretical results in \cite{1582743} were not compared with experimental evaluations of the gossip protocol. 
\revision{The scope of their paper differs from ours: it focuses on the correctness analysis, 
leaving out performance analysis of the proposed protocol.}

%
%

The paper is organized in the following sections. Sec.~\ref{sec:back} outlines the class of 
gossip protocols that can be analyzed with our framework, presented in Sec.~\ref{sec:method}. 
We consider two case studies, Newscast and Shuffle, and demonstrate the modelling and derivation of the state 
transition matrix for them, in Sec.~\ref{sec:newscast} and Sec.~\ref{sec:shuffle}, respectively. 
Sec.~\ref{sec:newscast} demonstrates the modelling of push and pull versions of Newscast. 
In Sec.~\ref{sec:lossy}, we model and analyze Shuffle in the presence of 
message loss. Sec.~\ref{sec:cyclon} 
stretches the scope of our theoretical framework to gossip-based membership protocols like Cyclon. 
Sec.~\ref{sec:conclusion} concludes the paper.

\section{Background}
\label{sec:back}
\begin{figure}[!ht]
\centering
\includegraphics[scale=0.95]{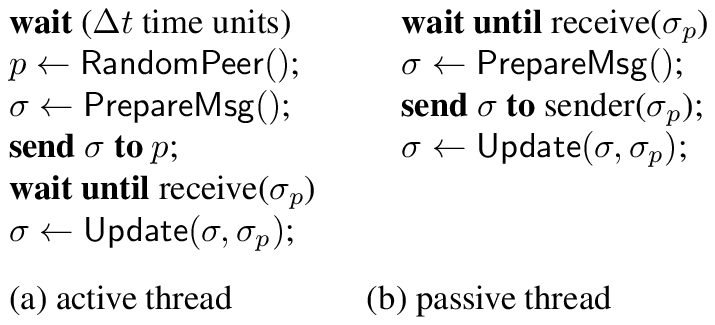}
\caption{Skeleton of a gossip protocol}
\label{fig:skeleton}
\vspace{-0.5cm}
\end{figure}

In this section, we specify a class of gossip protocols 
for which our modelling framework is applicable. We consider large networks, where nodes
interact in a peer-to-peer style.  All nodes have a common agreement on the frequency of gossiping. 
Each node stores local data in its \emph{cache} and executes two different threads, 
an active and a passive one (see Fig.~\ref{fig:skeleton}).  The active thread periodically 
initiates a contact with a randomly chosen peer $p$ by sending it a (sub)set $\msg$ of its cache, 
and waits for a reply. Upon reception of the reply ${\msg}_p$, the node  
updates its cache based on the cache, $\msg$ and ${\msg}_p$.

The passive thread waits for a message sent by the active thread of a neighbor and 
replies to it with its data $\msg$. The node then updates its cache based on the stored, 
received and sent data. The exchange message with its content is called an \emph{exchange buffer}.
Nodes that execute their active thread are \emph{initiators}, and nodes that execute their passive thread are 
called \emph{contacted}.

The generic protocol, described above, is a \emph{push-pull} gossip protocol. Other variations of gossip protocols 
are ones in which the initiator only sends local data to its gossip partner (\emph{push}), and ones in which an 
initiator only requests data from its peer (\emph{pull}). All three versions of gossip protocols are covered by our 
framework. 

The random peer selection ${\sf RandomPeer()}$ is based on the set of neighbors as 
determined by the underlying network graph. The nature of the data and the result of the generic operations in Fig.~\ref{fig:skeleton}
are application-dependent. In gossip-based information dissemination protocols like \cite{JKS03,GVS06}, 
a finite list of news items composes the local cache of a node. 
The operation $\msg \leftarrow {\sf PrepareMsg}()$ in Fig.~\ref{fig:skeleton} selects a 
random (or predefined) set of items from the cache of the node. The method ${\sf Update}$ merges the 
list of old items with the list of received items. Properties of the  
protocols that can be analyzed within our framework include the number of copies 
of an item in the network over time and the speed at which items spread throughout the network.
%

In gossip-based membership management protocols, such as \cite{VoulgarisGS05,1073871}, a 
cache of each node consists of a finite set of its peer IP addresses, \revision{so-called the \emph{partial view} 
of the membership of the network}.  
The method ${\sf Update}$ produces a sample of the union of the old and the received views. 
The performance metrics of these protocols include a distribution of the partial view 
size, and the number of nodes reached in the presence of node failures. 
We refer to \cite{1317388} for other applications of gossip protocols. 

As we demonstrate later in this paper, our framework models gossip-based information 
dissemination protocols, but \revision{it can be also applied} to other type of applications, 
such as gossip-based membership protocols.


\section{Pairwise Interaction Model}
\label{sec:method}

An exchange of data items between nodes is modelled as a state transition, capturing 
the presence or absence of an observed data item $\theitem$ before and after a gossip interaction between \emph{two nodes:  
an initiator $A$ and a contacted node $B$}. Each state is then a pair $(a,b)$ of bits, each indicating 
the presence (if equal to $1$) or the absence (if equal to $0$) of the item in the cache of $A$ and 
$B$, respectively. 

\begin{figure}[!t]
\centering
\includegraphics[scale=0.95]{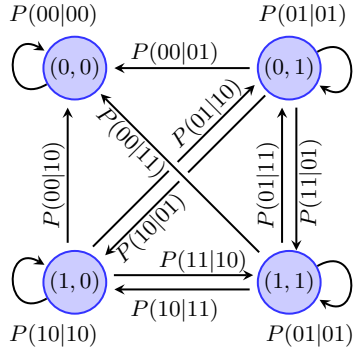}
\caption{Pairwise interaction model}
\label{fig:allsc}
\end{figure}

The model has four possible states of the caches of $A$ and $B$: (1) when both 
hold $\theitem$, (2--3) either of the caches holds the item $d$, and (4) neither cache holds 
$\theitem$. These correspond to the states $(0,0)$, $(0,1)$, $(1,0)$, and $(1,1)$ in 
the state transition diagram, shown as Fig.~\ref{fig:allsc}. 
Transitions from one state to another are labelled by the respective transition probabilities 
$P(a_2 b_2|a_1 b_1)$, where $a_1 b_1$ is the state before a gossip interaction, and  $a_2 b_2$ is the state 
after the interaction, with $a_i,b_i\in\{0,1\}$. $a_1$, $a_2$ and $b_1$, $b_2$ 
correspond to states of nodes $A$ and $B$. For instance, $P(01|10)$ means that node $A$ had 
$\theitem$ before the interaction, which it passed on to $B$, afterwards. 

The building blocks of all transition probabilities $P(a_2 b_2|a_1 b_1)$ of the state diagram are two  
probabilities: (i) the probability $\probsn$ of an item to be selected by a 
node from its cache for a gossip, and (ii) the probability $\probdn$ that an item is replaced by 
another one, received by its node in the gossip. The expressions for $\probsn$ and $\probdn$ depend on  
specifics of the operations ${\sf PrepareMsg}$ and ${\sf Update}$, respectively. 
Moreover, these probabilities are functions of the \emph{number of items} $\items$, \emph{exchange buffer size} 
$\buffer$, and \emph{cache size} $\cache$, and $0 < s \leq c$. We assume that $c < n$, \ie{} all $n$ items cannot be 
stored in a single local cache. 
We explain our framework by considering 
state transitions and computing the corresponding probabilities for several case studies. 

Our state diagram differs from a Markov chain of the system: it expresses a state of gossiping pair 
rather than state of all nodes. The transition probabilities are used with other frameworks, e.g., mean-field framework 
and numerical simulations, to study the emergent behaviour of 
system. 
\section{Case study: Newscast}
\label{sec:newscast}

We now briefly describe our first case study, a simple push-pull information propagation
protocol. It is a variation of the Newscast protocol \cite{newscast:tr}. In the original 
version of Newscast, each item is paired with a timestamp indicating when it was created. 
%
The original Newscast protocol can \revision{also be modelled} using our analytical 
framework; we will come back to the details in Sec.~\ref{sec:cyclon}.

The basic idea of the protocol is a periodic exchange of data items published by nodes in 
the network. Items can be, for example, a number or network address (IP 
and port) of a node. For now, we assume that nodes do not crash and communication 
channels are failure-free. 

\subsection{Specification}
The protocol operates in a wide area network\footnote{In the network of gossiping nodes, an information 
dissemination is fully dictated by gossiping frequency instead of communication latencies (cf.~\cite{newscast:tr}).}. 
A node periodically picks a random peer and exchanges $\buffer$ random items with it.
The node then selects $\cache$ random items for its new cache among the received items 
and those in its cache. The methods in Fig.~\ref{fig:skeleton} for Newscast are summarised 
in the following table:
\begin{center}
{\footnotesize
\begin{tabular}{|c|l|}
\hline
{\bf method} &  {\bf operation}\\
\hline
\oper{RandomPeer()} & select a peer uniformly at random\\
\hline
\oper{PrepareMsg()} & select $\buffer$ random items\\
\hline
\oper{Update($\msg,{\msg}_p$)} & store only $\cache$ random items, \\
                              & selected among the items in its cache \\
                              & and the received items ${\msg}_p$ \\
\hline
\end{tabular}
}
\end{center}


\subsection{Transition diagram}

We analyse the spread of a generic item, which we call $d$, using the framework  
introduced in Sec.~\ref{sec:method}. As stated previously, there are four possible 
states of the caches of $A$ and $B$. 

The state $(0,0)$ has only one outgoing transition, 
back to itself. $P(00|00) = 1$ since if $A$ and $B$ do not have $\theitem$ before 
their exchange then they clearly still do not have $\theitem$ after the exchange. 
We now determine values for other probabilities $P(a_2 b_2|a_1 b_1)$. 
Note that due to the symmetry of information exchange between gossiping 
nodes in the protocol, 
\(
P(a_2 b_2|a_1 b_1)=P(b_2 a_2|b_1 a_1);
\)
\revision{and the probabilities $\probdn$ and $\probsn$ are the same for both 
initiator and contacted node.} 
We use the following notations for complementary probabilities: $\probnotsn$ for $1-\probsn$, and 
$\probkn$ for $1-\probdn$. 

\begin{scenario}[$0,1$] Before gossip, $\theitem$ is only in the cache of node $B$.
\begin{compactdesc}
 \item[$a_2 b_2=01$:] $B$ neither sends nor drops $\theitem$ from its cache, or 
 $B$ sends and keeps $\theitem$, and $A$ drops it, \ie,\ the 
 probability is \(P(01|01)= \probnotsn \cdot \probkn + \probsn \cdot \probkn \cdot \probdn \).
 \item[$a_2 b_2=10$:] $B$ selects and drops $\theitem$, and $A$ keeps it; \ie, 
 the probability is \(P(10|01)=\probsn \cdot \probdn \cdot \probkn \).
 \item[$a_2 b_2=11$:] $B$ sends $\theitem$, and both nodes keep it; \ie, 
 \(P(11|01)=\probsn \cdot \probkn \cdot \probkn\).
 \item[$a_2 b_2=00$:] $B$ sends $\theitem$ and both nodes drop it, or $B$ does not send and drops $\theitem$; 
 \( P(00|01) = \probsn \cdot \probdn \cdot \probdn + \probnotsn \cdot \probdn \).
\end{compactdesc}
\end{scenario}
\begin{scenario}[$1,1$] Before gossip, $\theitem$ is in the caches of both nodes.
\begin{compactdesc}
 \item[$a_2 b_2=01$:] node $A$ drops $\theitem$ while updating its cache, but 
 $B$ keeps it; \ie, 
 \(P(01|11)= \probkn \cdot \probdn \)
 \item[$a_2 b_2=10$:] this transition is symmetric to the previous one.
 \item[$a_2 b_2=11$:] neither $A$ nor $B$ drops the item $\theitem$; \ie, \(P(11|11)= (\probkn)^{2} \).
\item[$a_2 b_2=00$:] both $A$ and $B$ drop $\theitem$; \ie, \(P(00|11) = (\probdn)^{2}\).
\end{compactdesc}
\end{scenario}

All transition probabilities are summarized in Fig.~\ref{fig:trNC}

\begin{figure}[!h]
\hrule
\vspace{-0.3cm}
{\scriptsize
\begin{minipage}[t]{0.48\linewidth}
\begin{align*}
P(01|01) &= P(10|10) = (\probnotsn + \probsn \cdot \probdn) \cdot \probkn \\
P(10|01) &= P(01|10) = \probsn \cdot \probdn \cdot \probkn \\
P(11|01) &= P(11|10) = \probsn \cdot \probkn \cdot \probkn \\
P(00|01) &= P(00|10) = (\probsn \cdot \probdn + \probnotsn) \cdot \probdn \\
P(01|11) &= P(10|11) = \probkn \cdot \probdn
\end{align*}
\end{minipage}

\begin{minipage}[t]{0.48\linewidth}
{\abovedisplayskip-1pt
\belowdisplayskip-1pt
\begin{align*}
P(11|11) &= {(\probkn)}^2  \hspace*{0.5cm} & P(00|11) &= {(\probdn)}^2 \hspace*{0.5cm} & P(00|00) &= 1 
\end{align*}}%
\end{minipage}}%
\hrule
\caption{Transition probabilities}
\label{fig:trNC}
\end{figure} 

\subsection{Building blocks: $\probsn$ and $\probdn$}
We now derive the expressions for the probabilities $\probsn$ and $\probdn$.
The following analysis assumes that all node caches are full; that is, 
the network is already running for a while.

\begin{figure}[!h]
        \centering                                             
        \includegraphics[scale=0.8]{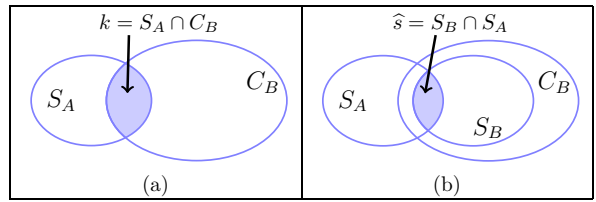}
        \caption{a) Items sent by $A$ that are in the cache of $B$; 
        b) Items in common for the exchange buffers of both $A$ and $B$. }
        \label{fig:kssets}
        \vspace{-0.25cm}
\end{figure}

Consider nodes $A$ and $B$ engaged in an exchange, and let $B$ receive the exchange 
buffer $S_A$ from $A$. Let $k$ be the number of duplicates (see Fig.~\ref{fig:kssets}), 
\ie,\ the items of an intersection of the node cache $C_B$ and the exchange buffer 
of its gossip partner $S_A$ (\ie,\ $S_A \cap C_B$). 

The probability of selecting an item $\theitem$ in the cache is the number of selected items 
$\buffer$ divided by the total number of items in the cache $\cache$: 
\( \probsn = \probs. \)
Thus, the probability that an item $\theitem$ in the cache is not 
selected is: \( \probnotsn = \frac{c-s}{c}. \)

Among the $\cache$ items in $C_B$, there are $k$ items also in $S_A$; thus, only 
$\cache$ random items of the total $c + s - k$ items in $C_B \cup (S_A \sminus C_B)$ 
can be kept: \( \probkn(k) = \frac{c}{c+s-k}. \)
Thus, the probability $\probdn$ that an item in $C_B \cup (S_A \sminus C_B)$ 
is dropped from $C_B$, given $k$ items in $S_A \cap C_B$ is as follows: 
\begin{align}
\label{eq:probdNC}
\probdn(k) = 1- \frac{c}{c+s-k}.
\end{align}
Assuming uniform sampling of items, the average value of $k$ is $s \cdot \frac{c}{n}$. Thus, the probability 
of dropping an item after the exchange $\probdn$ is $ \frac{1}{1+cn/(s(n-c))}$. 
Later 
we discuss a case when the assumption of uniform sampling does not hold.  

%
%


\subsection{Validation}

To validate our theoretical results, we simulated Newscast in a round-based fashion similar 
to simulations in PeerSim \cite{peersim}. A new item $\theitem$ is initially 
introduced in a network of $2500$ nodes at one node. In this fully connected network, caches of 
all nodes are full and uniformly populated by $\items=500$ items. After each \gp{}  
round, we measure the total number of copies of $\theitem$ in the network (\emph{replication} property), 
and how many nodes in total have seen $\theitem$ over time (\emph{coverage} property). 

\paragraph*{Simulations with the simplified Newscast} 
Each node in the network has a cache size of $\cache=100$. Once in a round, every node selects uniformly at random 
one node from the network of $2500$ nodes and exchanges $\buffer=50$ random items. To make a fair comparison with 
the simulations with the model, we let the nodes gossip for 1000 rounds items other than $\theitem$; items are 
replicated and the replicas fill the caches of all nodes. At the round 1000, the observed item $\theitem$ is inserted 
into the network at a random location. From that round on, replication and coverage are measured by the end of each 
round.  

\paragraph*{Simulations with the model} 
For the simulations with the model, the system parameters $\items, \cache$ and $\buffer$ are set to 500, 
100 and 50, respectively. Each node in the network only maintains a state variable which indicates the 
presence or absence of $\theitem$. Nodes update their state in pairs according to the transition 
probabilities in Fig.~\ref{fig:trNC}. While in the actual protocol, nodes update the contents of their caches, 
in the model simulations, nodes update only their state variables. Since we do not need a startup time 
for the simulations with the model, at the round 0, we set the state of a random node to 1, and  
all others to 0. From that round on, we track the states of the nodes. 

\begin{figure}[!t]
         \vspace{-0.25cm}
        \begin{minipage}[t]{0.49\linewidth}
         \centering\hspace*{-0.6cm}
        \includegraphics[scale=0.5]{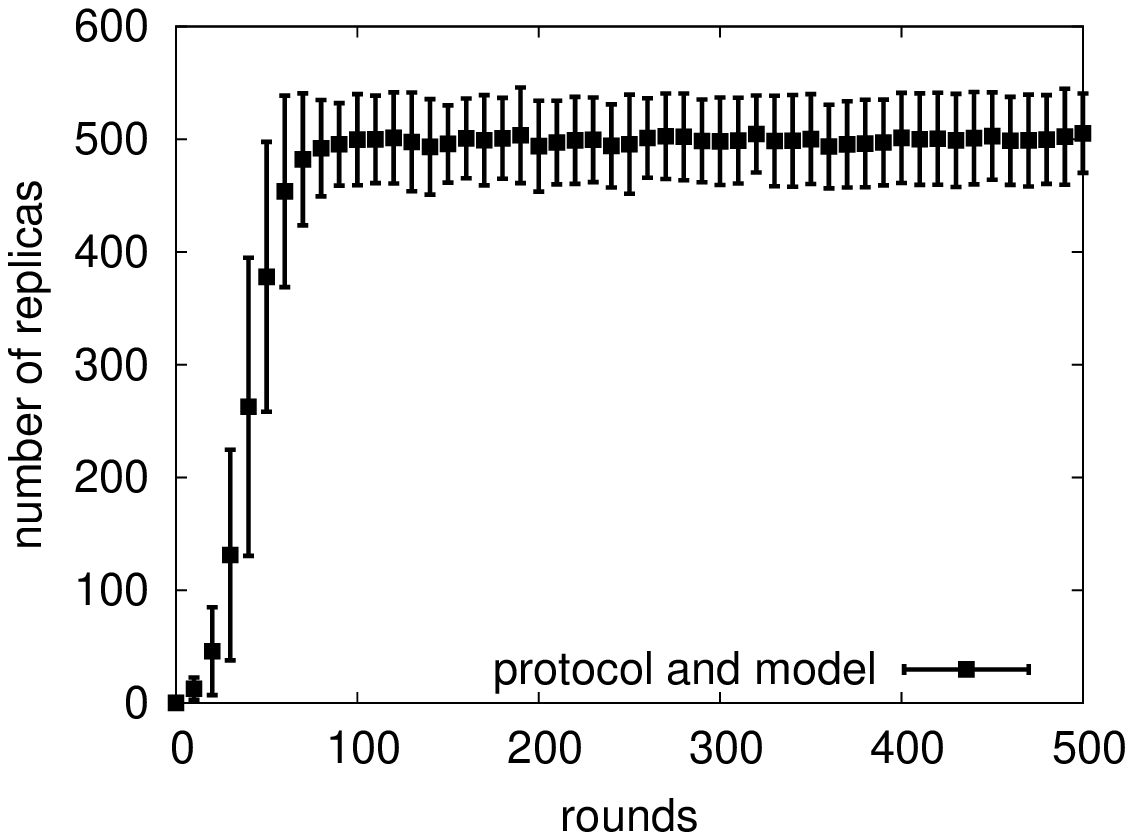}
        \end{minipage}
        \begin{minipage}[t]{0.49\linewidth}
         \centering\hspace*{-0.5cm}
        \includegraphics[scale=0.5]{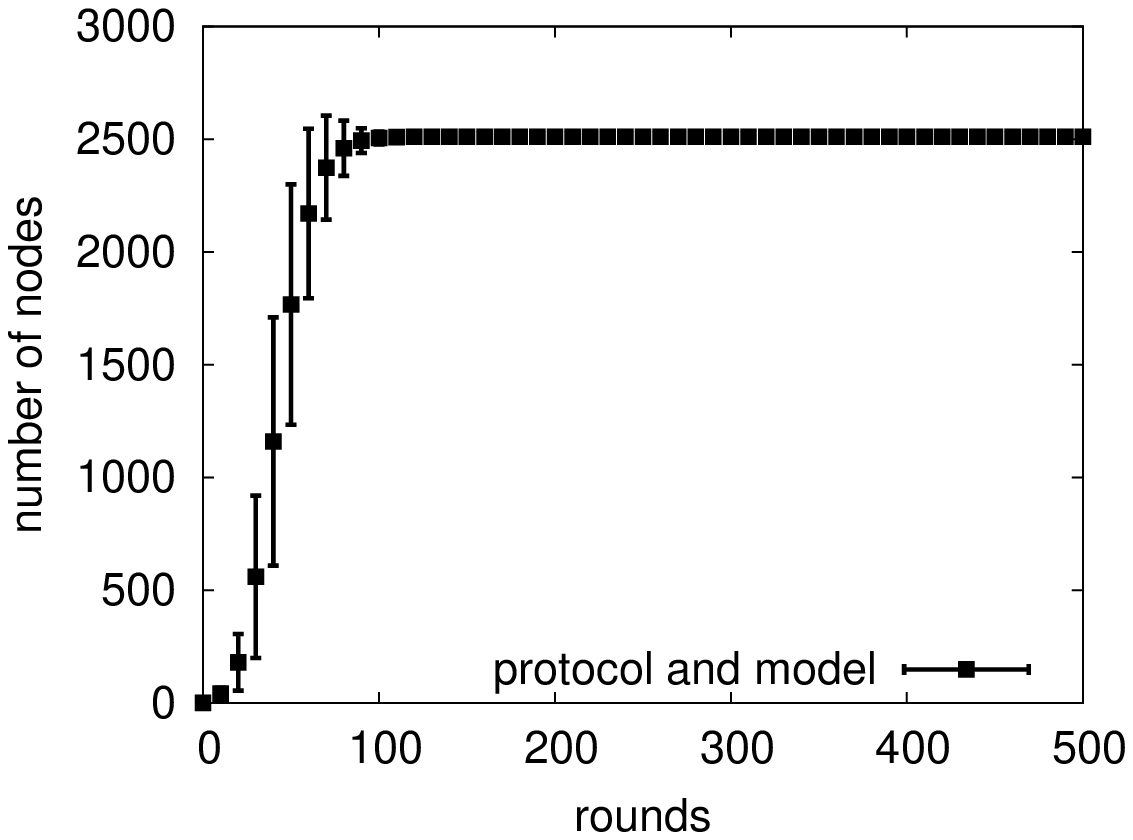} 
        \end{minipage}
        \caption{\textbf{Clique:} Replication (left) and coverage (right) of item 
        $\theitem$ for Newscast and its model, with $c = 100$, $n=500$, $s=50$.}
        \label{fig:propNewscast}
         \vspace{-0.25cm}
\end{figure} 

The graphs in Fig.~\ref{fig:propNewscast} depict the performance of the simplified Newscast 
and of the analytical model in terms of replication (left) and 
coverage (right) of $\theitem$. The results are collected from 1000 simulation runs of the protocol and 
1000 runs of the model. In approximately $72\%$ of the all runs, the observed item disappears. In 
the graphs, however, we only considered the runs where the item did not disappear; e.g., 
to obtain 1000 runs of the protocol displayed in Fig.~\ref{fig:propNewscast}, 
we performed a total of 3700 runs. Assuming a normal distribution of the results, the 
confidence interval for 95\% confidence is 16 times narrower than the standard deviations drawn on the graphs.


Fig.~\ref{fig:propNewscast} show the average and standard deviation of the successful 
runs with Newscast. The standard deviation bars on the graphs are the standard deviation of the replicas 
and number of nodes that have seen item in the system. 
For all successful runs, the network reaches the equilibrium, in which there are around 500 replicas 
of $\theitem$. Due to random gossiping, item $\theitem$ is eventually seen by all nodes in the network, 
when coverage reaches all $2500$ nodes. 

We have chosen this type of network graph for a basic introduction of our framework since a uniform distribution 
of items over the network is clearly a valid assumption in a fully connected network of nodes executing Newscast. 
We will turn to other types of network graphs in the next case study. 
Our framework allows for modelling of push-only as well pull-only gossip protocols, see Appendix~\ref{sec:onedir}.

\section{Case Study: Shuffle}
\label{sec:shuffle}

We now move on to another case, the Shuffle protocol introduced in \cite{GVS06} and 
originally analyzed in \cite{BGFSCN09} using our framework. This section is intended  
to make the reader familiar with the analysis of this protocol. 

\subsection{Specification}
Shuffle aims at the conservation of data in an ad hoc network. 
Each node maintains a cache of data items that are disseminated throughout 
the network. The dissemination is done by periodic exchange of $\buffer$ 
random items between two gossiping peers. 

The procedure of selection of items is similar to the one in 
Newscast, but items are discarded according to a different 
policy: (i) a node cannot discard an item unless it is sent to the gossip partner;
(ii) the partner is not allowed to drop received items.

A node can keep only $\cache$ items in its cache after the exchange.
A peer will favor dropping items it has just sent over other items 
in its cache. Note that in this way, an exchanged item will always 
be preserved, and possibly even replicated. 
Since the peer, in its turn, has agreed to store received items in its cache, 
discarding items does not lead to the loss of information in the network 
(if there is no message loss). 
The generic routines in Fig.~\ref{fig:skeleton} can be summarized for Shuffle as 
follows:

\begin{center}
{\footnotesize
\begin{tabular}{|c|l|}
\hline
{\bf method} &  {\bf operation}\\
\hline
{\sf\scriptsize RandomPeer()} & select a peer uniformly at random\\
\hline
{\sf\scriptsize PrepareMsg()} & select $\buffer$ random items\\
\hline
{\sf\scriptsize Update($\msg,{\msg}_p$)} & $\bullet$ add ${\msg}_p$ received entries to the cache;\\
                                    & $\bullet$ remove duplicated items; \\
                                    & $\bullet$ remove items among $\msg \setminus {\msg}_p$ uniformly \\
                                    & at random until the cache has $\cache$ items.\\
\hline
\end{tabular} 
}
\end{center}

\subsection{Transition Diagram}

Unlike in Newscast, nodes are not allowed to discard 
the items received from their gossip partners. Taking this difference into 
account, the transition probabilities for Shuffle are quite easy to derive 
from the transitions in Fig.~\ref{fig:trNC}. Note that for this protocol the state $(0,0)$ has only a 
self-transition, and no other outgoing or incoming transitions, because of the 
preservation nature of the protocol. That is,  
%
if either nodes send an item, its partner keeps this copy as well, and if an item 
is not among the selected for a shuffle, the item is not replaced by another one. 
%
%
The transition probabilities can be easily computed, see Fig.~\ref{fig:tranSh}. 
For more details on derivation of these transition probabilities, we refer to \cite{BGFSCN09}. 

\begin{figure}[!h]
\hrule
\vspace{-0.3cm}
{\footnotesize
\begin{minipage}[t]{0.48\linewidth}
 \begin{align*}
 P(01|01) &= P(10|10) = \probnotsn \hspace*{-0.1cm} & P(10|01) &= P(01|10) = \probsn \cdot \probdn
 \end{align*}
 \end{minipage}
 \begin{minipage}[t]{0.48\linewidth}
 {\abovedisplayskip0pt
 \belowdisplayskip-1pt
 \begin{align*}
 P(01|11) &= P(10|11) = \probsn \cdot \probnotsn \cdot \probdn &P(00|00) &= 1 \\
 P(11|10) &= P(11|01) = \probsn \cdot \probkn \\
 P(11|11) &= 1- 2 \cdot \probsn \cdot \probnotsn \cdot \probdn 
 \end{align*}}%
\end{minipage}}
\hrule
\caption{Transition probabilities for Shuffle}
\label{fig:tranSh}
\end{figure}

$\probsn$ remains the same as for Newscast, \(\probsn = \frac{s}{c}\). However, $\probdn$ 
is different for Shuffle, and we derive it in the next section. 

\subsection{Probability of Dropping an item}
\label{sec:shuffle:pdrop}

$\probdn$ represents the probability that an item that can be overwritten is 
indeed overwritten by an item received by its node in the exchange. 

$\probdn$ depends on the number of items both gossiping nodes have in common, 
in particular, how many of such items the contacted node receives during the exchange. 
To derive the expression for $\probdn$, we assume a uniform distribution of items over 
the network; in the absence of message loss, this assumption is supported by experiments 
in \cite{GVS06,JGKS04,BGFSCN09} and analysis in \cite{pdcat:09}. 

Consider again Fig.~\ref{fig:kssets}. 
By assumption, $C_A$ and $C_B$ are a uniform selection from the entire population 
of $\items$ items. $S_A$ and $S_B$ are chosen uniformly at random from $C_A$ and $C_B$, 
respectively. 
After the exchange, node $B$ has to allocate items from $S_A$ in its cache $C_B$, 
taking into the account the duplicated items in $S_A \cup C_B$ and $S_A \cup S_B$.  
Basically, the items in $S_B$ that are not in $S_A$ are replaced by items in 
$S_A$ that are not in $C_B$. So, every item from $S_A \setminus S_B$ that is in 
$C_B$ means that one item from $S_B \setminus S_A$ can be kept in $C_B$. In other words, 
the probability $\probdn$ that an item from $S_B \cup S_A$ is dropped from $C_B$ 
equals to the probability that an item from $S_A \setminus S_B$ is not in $C_B$. 
By uniform distribution, the probability is $\frac{n-c}{n-s}$. 
Thus, \( \probdn \approx \frac{n-c}{n-s}.\) 

This approximation of $\probdn$ for Shuffle has been 
successfully established through experiments \cite{BGFSCN09}, and used for modelling 
and optimization in \cite{BGFSCN09,BF09}.


%
%

\section{Shuffling through Lossy Channels}
\label{sec:lossy}

So far, the analysis has relied on the assumption that nodes interact in a  
perfect, lossless communication environment. We now drop this 
assumption, and consider ad hoc networks where nodes are continually 
communicating with each other over an unreliable medium.
We now explore the impact of lossy channels on a \gp{} protocol with a 
push-pull information exchange. 
%

\subsection{Assumptions}
  Every message sent now has a fixed, positive probability to be lost due to a 
  disturbance of the communication medium. 
 As explained in the introduction, 
  we no longer assume that the shuffle procedure is atomic. 

 \begin{figure}[!b]
  \vspace{-0.5cm}
  \centering
  \includegraphics[scale=0.9]{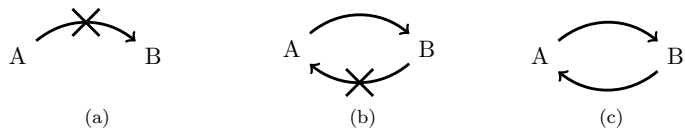}
  \caption{Scenarios of communication with lossy channels: (a) loss of the request message, 
  (b) loss of the reply message, (c) gossip without loss.}
  \label{fig:comm}
\end{figure}
  
  There are three general cases in pairwise 
  communication with respect to message delivery, as depicted in Fig.~\ref{fig:comm}: 
  (a) node $A$ initiates a gossip with $B$ by sending a message, but the message is lost, 
  (b) node $A$ successfully initiates a gossip with $B$, but 
  a message from $B$ is lost on its way to $A$, and 
  (c) a gossiping pair receives messages from each other. 

\subsection{Transitions}
We take the analytical model of Shuffle described in Sec.~\ref{sec:shuffle} as  
starting point. In our current model, 
the state $(0,0)$ is no longer isolated, since there is a possibility to remove 
the only copy of $\theitem$ from the cache of the node $B$, in the scenario shown in Fig.~\ref{fig:comm}(b).

We again express the transition probabilities $P(a_2 b_2|a_1 b_1)$ in terms of $\probsn$, $\probdn$. 
To analyze the protocol in the presence of message loss, we introduce into the formal model 
an additional input parameter, the probability $\probls$ that a message is not 
delivered to a node due to channel loss. We use $\probnls$ for $1 - \probls$. 

\begin{scenario}[$0,0$] Before shuffle, neither $A$ nor $B$ have $\theitem$ in their cache.
\begin{compactdesc}
 \item[$a_2 b_2=00$:] neither $A$ nor $B$ have item $\theitem$ after a shuffle, because neither 
of them had it in the caches before the shuffle: $P(00|00)=1$.
\item[$a_2 b_2 \in \{01,10,11\}$:] cannot occur, because none of the nodes have item $\theitem$.
\end{compactdesc} 
\end{scenario}
\begin{scenario}[$1,0$] Before shuffle, $\theitem$ is only in the cache of node $A$.
\begin{compactdesc}
 \item[$a_2 b_2=01$:] occurs only when both the request of $A$ and the reply of $B$ are successful and node 
$A$ selects and drops $\theitem$:  
\(P(01|10)= {(\probnls)}^2 \cdot \probsn \cdot \probdn.\)
  \item[$a_2 b_2=10$:] $B$ does not have $\theitem$ because:
 (a) $A$ did not select $\theitem$, or (b) $A$ selected $\theitem$ but the request 
  message got lost on the way to $B$:  
\(P(10|10)= \probnotsn + \probls \cdot \probsn.\)
 \item[$a_2 b_2=11$:] both nodes $A$ and $B$ have a copy of $\theitem$ because: 
 (a) either both nodes received the gossip messages and $A$ selected $\theitem$ and kept it,
 or (b) $A$ selected $\theitem$ and the reply message from $B$ got lost; that is, 
\(P(11|10)= {(\probnls)}^2 \cdot \probsn \cdot \probkn + \probnls \cdot \probls \cdot \probsn \)
 \item[$a_2 b_2=00$:] cannot occur, as $A$ would only drop $\theitem$ if it 
 received a reply, which implies that $B$ would keep $\theitem$. 
\end{compactdesc}
\end{scenario}
\begin{scenario}[$0,1$] Before shuffle, a copy of $\theitem$ is only in the cache of node $B$.
\begin{compactdesc}
 \item[$a_2 b_2=01$:] only $B$ has $\theitem$ because the message from $A$ got lost 
 or, $B$ received the message, and: (a) it did not select $\theitem$ or (b) $B$ 
 selected $\theitem$, kept it, and reply got lost; i.e. the probability is 
\(P(01|01)= \probnls \cdot (\probnotsn + \probls \cdot \probsn \cdot \probkn) + \probls.\)
 \item[$a_2 b_2=10$:] both request and reply messages were successfully delivered and $B$ selected 
 and dropped $\theitem$, which amounts to the probability 
\( P(10|01)= {(\probnls)}^2 \cdot \probsn \cdot \probdn. \)
\item[$a_2 b_2=11$:] both messages were delivered successfully, and $B$ selected and kept $\theitem$; that is, 
$P(11|01)= {(\probnls)}^2 \cdot \probsn \cdot \probkn$.
 \item[$a_2 b_2=00$:] neither of the nodes have $\theitem$, because $B$ selected and dropped $\theitem$, but $A$ 
 did not receive the reply message. $P(00|01)= \probnls \cdot \probls \cdot \probsn \cdot \probdn$. 
\end{compactdesc}
\end{scenario}
\begin{scenario}[$1,1$] Before shuffle, $\theitem$ is in the caches of $A$ and $B$.
\begin{compactdesc}
 \item[$a_2 b_2=01$:] only $B$ has $\theitem$ since both messages were successfully received, 
 $A$ selected and dropped $\theitem$ while $B$ did not select it: 
 \(P(01|11)= {(\probnls)}^2 \cdot \probsn \cdot \probdn \cdot \probnotsn. \)
  \item[$a_2 b_2=10$:] only node $A$ has $\theitem$ because node $B$ selected $\theitem$, dropped it, and node $A$ 
 did not select $\theitem$: 
 $P(10|11) = \probnls \cdot \probnotsn \cdot \probsn \cdot \probdn$.
  \item[$a_2 b_2=11$:] after the shuffle both nodes have $\theitem$, because:
  \item[a)] the message from $A$ did not arrive at $B$, i.e. \(\probls\); or
  \item[b)] the message from $A$ was successfully received by $B$, but the reply message got lost, and:
   \begin{compactitem}[$\star$]
     \item $A$ did not select $\theitem$, while $B$ (i) did not select $\theitem$ as well, or (ii) selected and kept $\theitem$. 
     \(\probnls \cdot \probnotsn \cdot \probls \cdot (\probnotsn + \probsn \cdot \probkn);\) and
     \item node $A$ selected $\theitem$:
     \(\probnls \cdot \probsn \cdot \probls;\) or
   \end{compactitem}
  \item[c)] both nodes received each other messages, and:
    \begin{compactitem}[$\star$]
     \item $A$ did not select $\theitem$, while $B$ (i) also did not select it, or (ii) selected and kept it:
      \({(\probnls)}^2 \cdot \probnotsn \cdot (\probnotsn + \probsn \cdot \probkn);\) and 
     \item (i) $A$ selected and kept $\theitem$, while $B$ did not select $\theitem$, or 
     (ii) both nodes selected $\theitem$.
     \({(\probnls)}^2 \cdot \probsn \cdot (\probnotsn \cdot \probkn + \probsn).\)
    \end{compactitem}
 \hspace*{-0.75cm} Hence, $P(11|11) = 1-(2-\probls \cdot (3-\probls)) \cdot \probsn \cdot \probnotsn \cdot \probdn$.
\item[$a_2 b_2=00$:] discarding of an item by both nodes cannot occur.
\end{compactdesc}
\end{scenario}

While $\probls$ expresses the reliability of the communication channels,  $\probsn$ and 
$\probdn$ are derived based on the behavior of the protocol. $\probsn$ is again 
the random selection of items from the cache; an item has $\frac{\buffer}{\cache}$ 
chance of being selected, regardless of message loss. The calculation of $\probdn$ is 
complex, and we discuss it in the remainder of this section.

\subsection{The Probability of Dropping an Item}
\label{sec:general}
We now analyze how message loss affects one of the building blocks of 
our model: $\probdn$. In Sec.~\ref{sec:shuffle:pdrop},  
we have already found an expression for $\probdn$ in the case of Shuffle, relying on the assumption of a uniform 
distribution of items. Here, we revisit $\probdn$ and derive a general formula 
without making any assumptions about the distribution of items. Later on, we will explore 
how message loss affects the distribution of items, and will determine that it is the coupling 
of message loss and the topology of the network that affects $\probdn$. Having 
isolated the component of $\probdn$ that is affected by the message loss/topology 
coupling, we will propose the use of statistical data to calculate that specific part 
of the $\probdn$ expression.

\begin{figure}[!b]
        \vspace{-0.5cm}
        \centering                                             
        \includegraphics[scale=0.8]{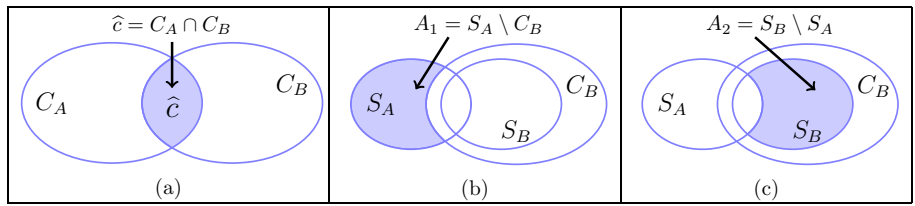}
        \caption{a) Items in common for the caches of both nodes $A$ and $B$; 
        b) Items sent by $A$ that are not in the cache of $B$; 
        c) Items sent by $B$ that are not in the exchange buffer of $A$.}
        \label{fig:msets}
\end{figure}

When a node selects an item to be sent to a neighbor, there is a probability 
$\probdn$ that the item will be dropped from the node's cache after the gossip 
exchange. The selected item may be dropped only when there is a need to create 
space for an item received from a gossip partner. Therefore, the probability 
$\probdn$ depends on the relationship between a) the new items received from the 
gossip partner (for which the node needs space, the shaded area in Fig.~\ref{fig:msets}(b),  
referred to as $A_1$), and b) the items that the node has selected to send to the 
gossip partner and is allowed to discard (the shaded area in Fig.~\ref{fig:msets}(c), 
referred to as $A_2$). In order to 
find an expression for $\probdn$, we need to calculate the probability of an 
item being in $A_1$ and the probability of an item being in $A_2$, which we 
will denote as $P(A_1)$ and $P(A_2)$, respectively.

Given two nodes, $A$ and $B$, that engage in a gossip exchange, we can represent 
their caches as sets $C_A$ and $C_B$, and the sets of items they exchange with each
other by $S_A$ and $S_B$, respectively.  The sets $C_A$ and $C_B$ may have common 
elements (items), see Fig.~\ref{fig:msets}(a). We define $\proboch$ as the probability 
of any item found in one of the caches to be also found in the other. Knowing $\proboch$, we 
can formulate $P(A_1)$ as $\probsn \cdot \left( 1-\proboch \right)$ and 
$P(A_2)$ as $\probsn \cdot \left( 1-\probsn \cdot \proboch \right)$. 
The probability $\probdn$ can then be expressed as:
{\abovedisplayskip5pt
 \belowdisplayskip5pt
\begin{align}
\probdn = \frac{P(A_1)}{P(A_2)} = \frac{1-\proboch}{1-\probsn \cdot \proboch}
\label{eq:probdn}
\end{align}}%
In order to calculate a value for $\probdn$, 
it is necessary to have a value for the probability $\proboch$. For the case of 
$C_A$ and $C_B$ being random samples of a population of $\items$ items, that is, 
under the assumption that items are uniformly distributed over the network we can analytically deduce that 
$\proboch = \frac{\cache}{\items}$. For networks with perfect communication channels,  
repeated execution of Shuffle results in a uniform distribution of items. 
In the next section, we look at the effect that lossy channels have on the 
distribution of items. 

\subsection{Uniform Distribution of Items: Does it Still Hold?}
As already mentioned, the calculation of $\probdn$ 
in the presence of message loss requires us to speculate about the contents of the gossip 
partner's cache. Assuming that the items in the caches of both gossip partners 
are random samples of the totality of items in the network, we can easily 
estimate $\proboch$ as $\frac{\cache}{\items}$ and, therefore, have an analytical expression 
for $\probdn \approx \frac{n-c}{n-s}$. In this section, we verify whether the uniform 
distribution of items, observed under no message loss, is still a valid 
assumption. 

\subsection*{The Importance of Randomness}
\begin{figure}[!ht]
        \centering
       \includegraphics[scale=0.8]{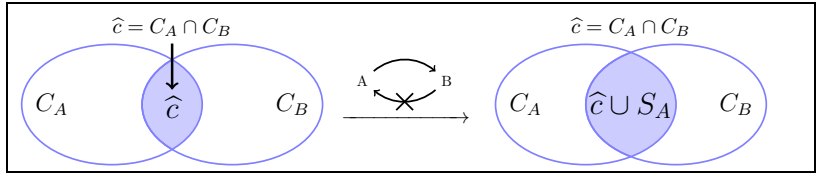}
       \caption{Correlation of the caches increases due to message loss}
       \label{fig:corr}
\end{figure}

To understand the effect of message loss on abstract level, 
consider one node in the network, executing Shuffle.  
We examine the scenario from Fig.~\ref{fig:comm}(b), where a reply message from 
a node to the initiating node is lost due to channel failure. 
When two nodes $A$ and $B$ gossip their local items to each other, the probability that 
the message from $B$ to $A$ is lost, is $\probnls \cdot \probls$. 
If the message of $B$ is not delivered to $A$, items $\chat \cup S_A$ will be common 
to cache $C_A$ and cache $C_B$ after the shuffle, since $B$, unaware of the failure, purges the 
sent items $S_B$, and $A$, which received no reply, keeps its $S_A$ items (see Fig.~\ref{fig:corr}). 

Over time neighboring nodes have a growing intersection of items in their cache.
If a node has a small neighborhood, the random sample becomes more biased towards the 
collection of items left in the neighborhood. Limited by the access to the storage 
space of the neighborhood, a set of neighbors maintains only a subset of all items. 
\revision{In general, the stronger the correlation between the caches content of the neighbours,  
the slower the spread of observed item throughout the network.}

For the fully connected graph, since it is the entire collection of the items every node has 
access to, the uniform distribution assumption in general remains. A larger neighborhood reduces the 
probability of repeated communication between two nodes, and due to the increased communication range, there is a faster 
exchange between distant areas of the network. 

\subsection*{Experimental Observations}
To support our claims that message loss affects the distribution of items, we conducted 
simulation experiments. Each simulation experiment has a startup 
period of 1000 rounds, during which $N=2500$ nodes gossip $\items=500$ different items under 
no message loss. We use three different network topologies: a fully connected network, 
a square grid with each internal node having 4 neighbors, and a network where every node has 4 randomly 
chosen neighbors (outdegree 4). By the end of the startup period, the items have been replicated, 
achieving uniform distribution. That is, every item has a probability 
$\frac{\cache}{\items} = \frac{100}{500} = 0.2$ of being present in a given node's 
cache. After the startup period has finished, the communication channel is 
set up to fail with a probability $\probls$.  We calculate, per round, the probabilities 
$P(11)$, $P(10)$ and $P(01)$ 
that an exchange involves both nodes having item $x$, only the initiator having item 
$x$, or only the gossip partner having item $x$, respectively.

\begin{figure}[!h]
       \vspace{-0.5cm}
        \centering
        \hspace*{-0.3cm}
        \begin{tabular}{@{}c@{}c}
        \hspace{-0.8cm}
       \includegraphics[scale=0.55]{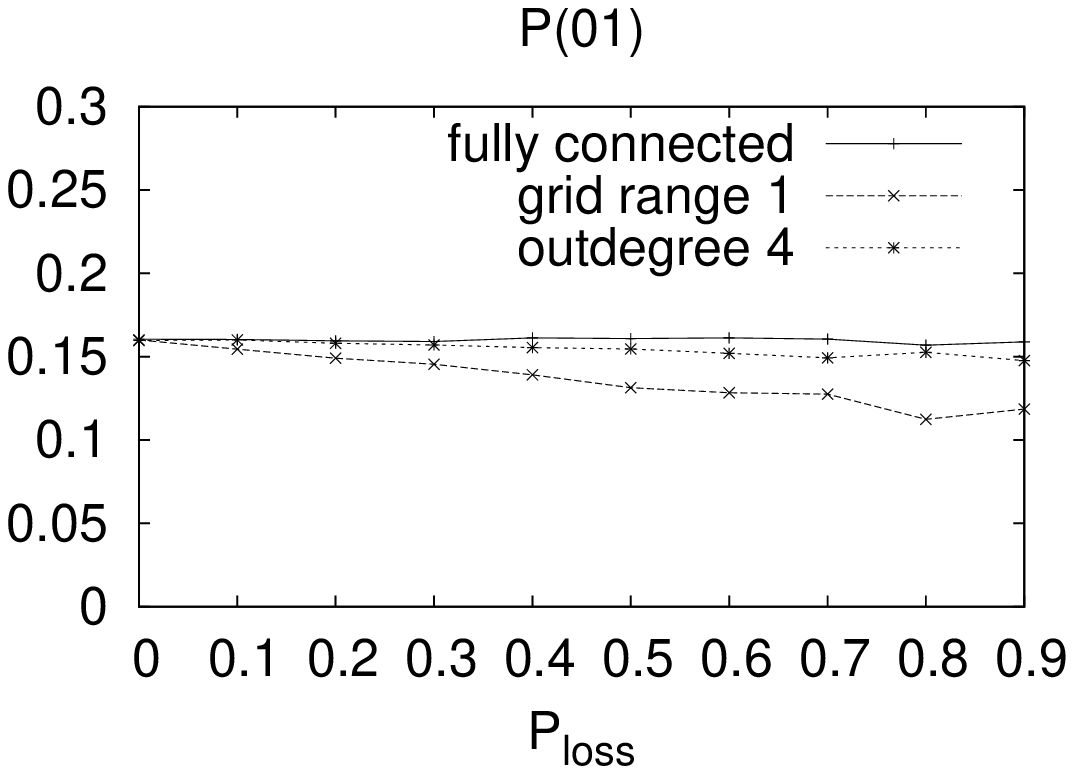}
       \hspace*{-0.9cm}
        & \includegraphics[scale=0.55]{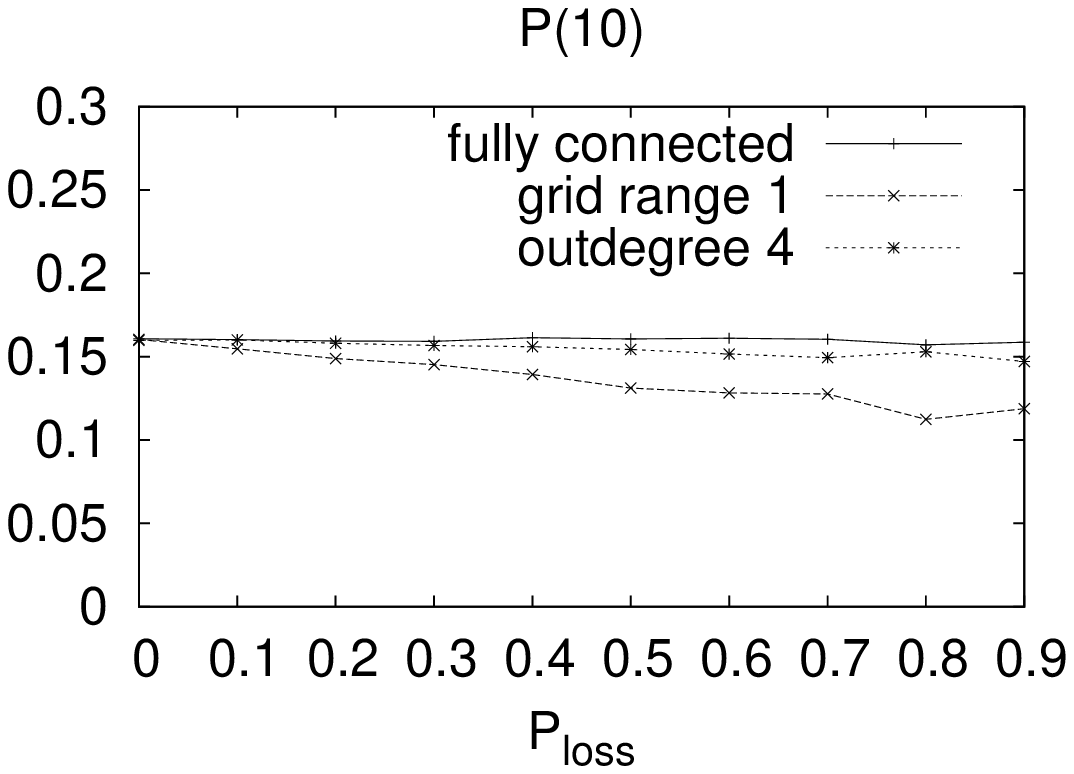} \\[-2ex]
        {\bf (a)} & {\bf (b)} \\[-2ex]
        \multicolumn{2}{c}{
        \includegraphics[scale=0.55]{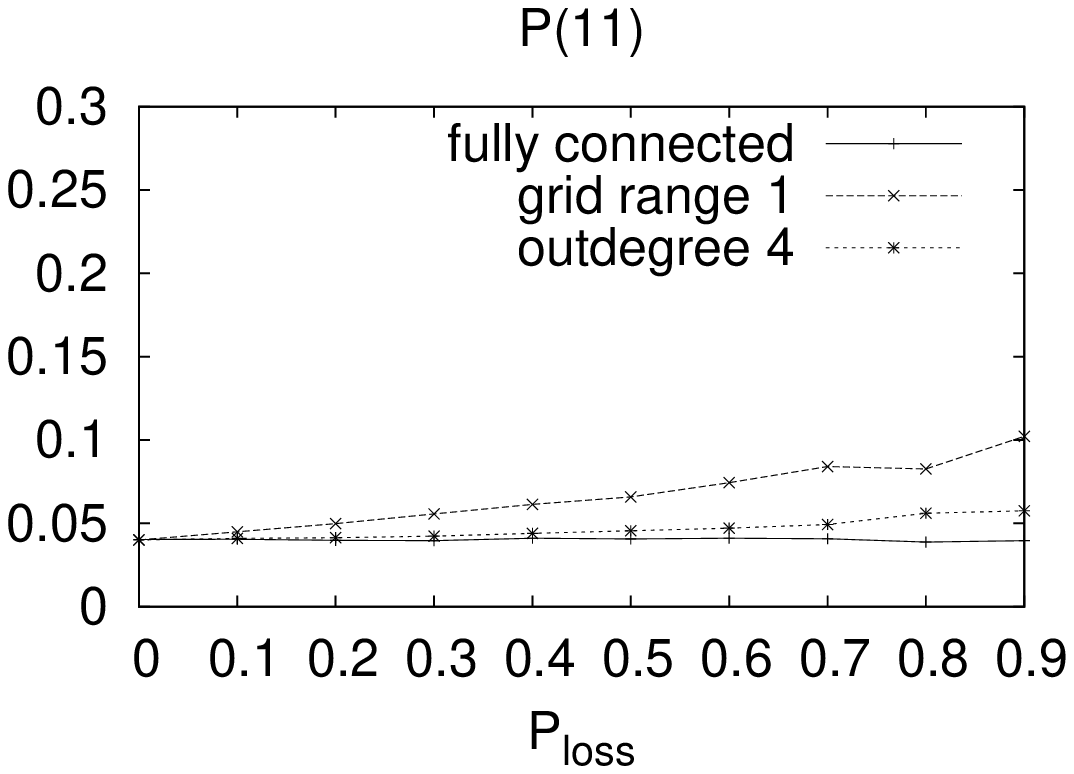}} \\[-2ex]
        \multicolumn{2}{c}{{\bf (c)}}
        \end{tabular}
        \caption{Probability that a gossip exchange occurs between a) a node 
        that does not have item $x$ and a node that has the item, b) a node 
        that has item $x$ and one that does not, and c) two nodes have item 
        $x$, for different topologies.}
        \label{fig:diffTopologies}
\end{figure}


Fig.~\ref{fig:diffTopologies} shows the average values of $P(11)$, $P(10)$ 
and $P(01)$ over 1000 rounds for different values of $\probls$ using the three 
topologies mentioned earlier. As expected, for the case of no message loss 
($\probls = 0$) the probabilities suggest a uniform distribution of items in 
the network. We can analytically deduce their expected value, which matches 
the experimental results, with 
$P(11)$ as $\frac{\cache}{\items} \cdot \frac{\cache}{\items} = 0.04$ and 
$P(10) = P(01)$ as $\frac{\cache}{\items} \cdot (1-\frac{\cache}{\items}) = 0.16$. 
However, as the probability of message loss increases, we observe some changes 
in $P(11)$, $P(10)$ and $P(01)$. For the case of the fully connected network, 
the probabilities remain stable, suggesting that the uniform distribution of 
items remains unaffected by message loss. With the other topologies, however, 
as message loss increases. the number of gossip interactions between nodes 
that both have item $x$ also rises. Since item $x$ is representative of all items 
in the network, the graphs suggest that, as a consequence of message loss, 
gossip partners have more items in common than they would have if items 
were uniformly distributed. In other words, due to message loss, a node is more 
likely to have items in common with a neighbor than with another random node 
in the network. Hence, the topology of the underlying network now plays a role 
in calculating $\probdn$.

\subsection{Calculating $\probdn$ based on Statistical Data}
\label{subsec:calculatingPdrop}

\begin{figure}[!h]
         \vspace{-0.5cm}
        \begin{minipage}[t]{0.45\linewidth}
        \centering\hspace*{-0.8cm}
        \includegraphics[scale=0.55]{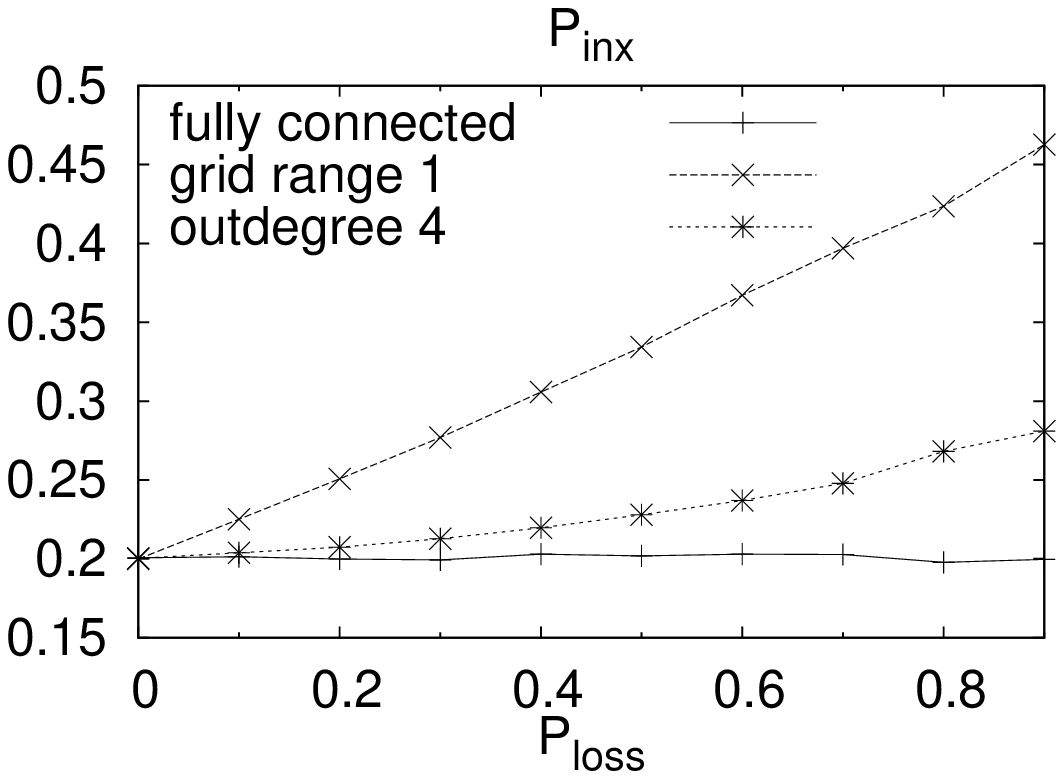}
        \end{minipage}
        \begin{minipage}[t]{0.45\linewidth}
        \centering\hspace*{-0.25cm}
        \includegraphics[scale=0.55]{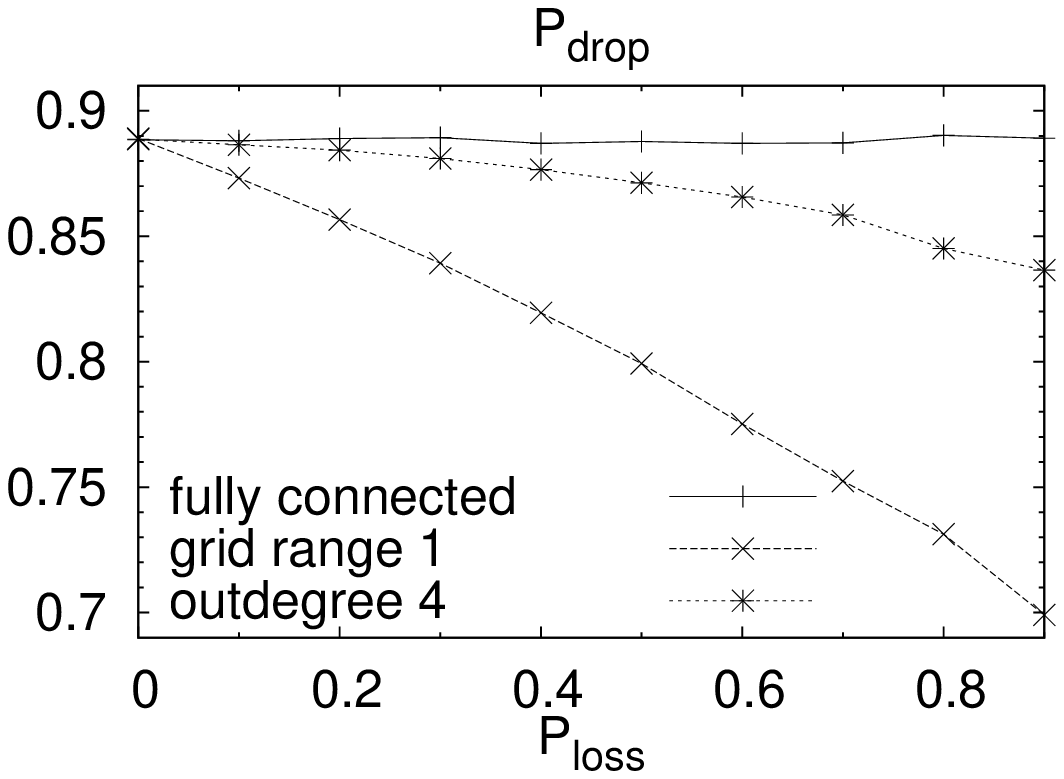} 
        \end{minipage}
        \vspace{-0.5cm}
        \caption{Probabilities $\proboch$ (left) and $\probdn$ (right), for different topologies. }
        \label{fig:p1given1}
\end{figure}

By now, we have established that in case of message loss
the structure of neighborhoods 
plays an important role in determining the distribution of items, which, in turn, determines the 
probability of an item found in a node's cache to be also found in the gossip partner's cache,  
$\proboch$. Our calculation of $\probdn$ depends on finding a value for $\proboch$. This topology 
varying component can be modelled analytically or measured experimentally from a single run of the protocol, 
for a given topology. 
%
%
Here, we opt for obtaining $\proboch$ for a given network graph from statistical data collected from experiments. With $\proboch$, 
we can proceed to calculate $\probdn$, obtaining the final building block of the model needed for 
validation.
We can calculate $\proboch$ from the probabilities $P(11)$, $P(10)$ and $P(01)$ measured 
experimentally in the previous section:
 {\abovedisplayskip5pt
  \belowdisplayskip5pt
\begin{align*}
\proboch = \frac{P(11)}{P(10)+P(11)}
\end{align*}}%
The left graph of Fig.~\ref{fig:p1given1} shows the values for $\proboch$ calculated for each experiment using 
a different $\probls$. Based on these values, we compute $\probdn$ using \eqref{eq:probdn}, as seen 
in right graph of Fig.~\ref{fig:p1given1}. As expected, the calculated values show that, in the face of message 
loss, different topologies yield different probabilities of dropping an item. $\probdn$ drops 
more harshly in the more clustered topologies, which suffer more from neighboring nodes having similar 
items as message loss increases. 

\subsection{Experimental Evaluation}
\label{subsec:expeval}
The experiments in this section simulate the case where a new item $\theitem$ is introduced by 
one node in the network in which all caches are full and uniformly populated by $\items=500$ 
items. Each experiment takes as input the topology of the network to determine which pairs 
of nodes can gossip. In all cases, we use a network of $N=2500$ nodes arranged in either 
of the three topologies mentioned in the previous section (a fully connected network, 
a square grid or a graph where every node has 4 randomly chosen neighbors). In the experiments 
that follow, after each \gp{}  
round, we measure the total number of occurrences of $\theitem$ in the network (replication), and how 
many nodes in total have seen $\theitem$ (coverage).

\paragraph*{Simulations with Shuffle} Each node in the network has a cache size of 
$\cache=100$, and sends $\buffer=50$ items when gossiping. In each round, every node randomly selects one 
of its neighbors and shuffles. In order to make a fair comparison with the simulations with 
the model, we let the nodes gossip for 1000 rounds with $\probls = 0$, to ensure that none of 
the $\items=500$ initial items is lost in the startup period before initiating the measurements 
of the properties. After this startup period of 1000 rounds, items are replicated and the 
replicas fill the caches of all nodes. At round 1000, $\probls$ is set to the desired value 
and item $\theitem$ is inserted into the network at a random location. From that moment on, we 
track its replication and coverage.


\begin{figure}[!b]
         \vspace{-0.75cm}
        \begin{minipage}[t]{0.49\linewidth}
        \centering\hspace*{-0.6cm}
        \includegraphics[scale=0.5]{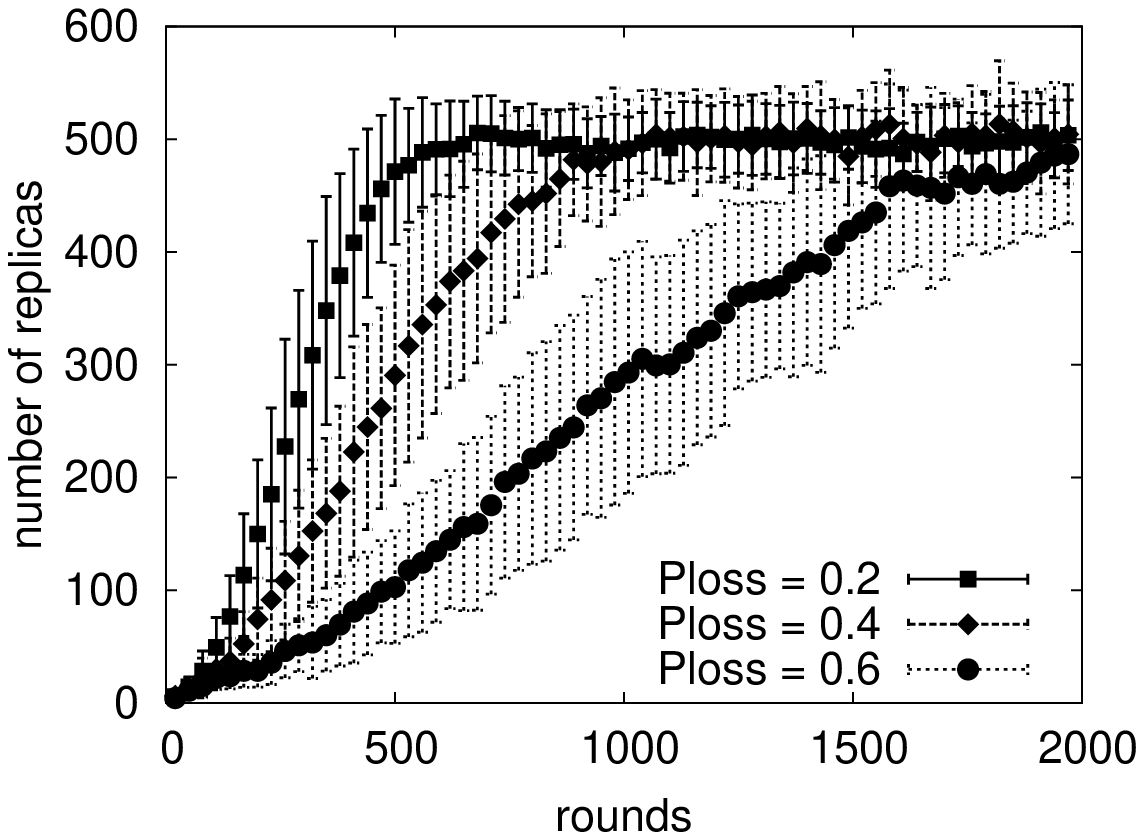}
        \end{minipage}
        \begin{minipage}[t]{0.49\linewidth}
        \centering\hspace*{-0.7cm}
        \includegraphics[scale=0.5]{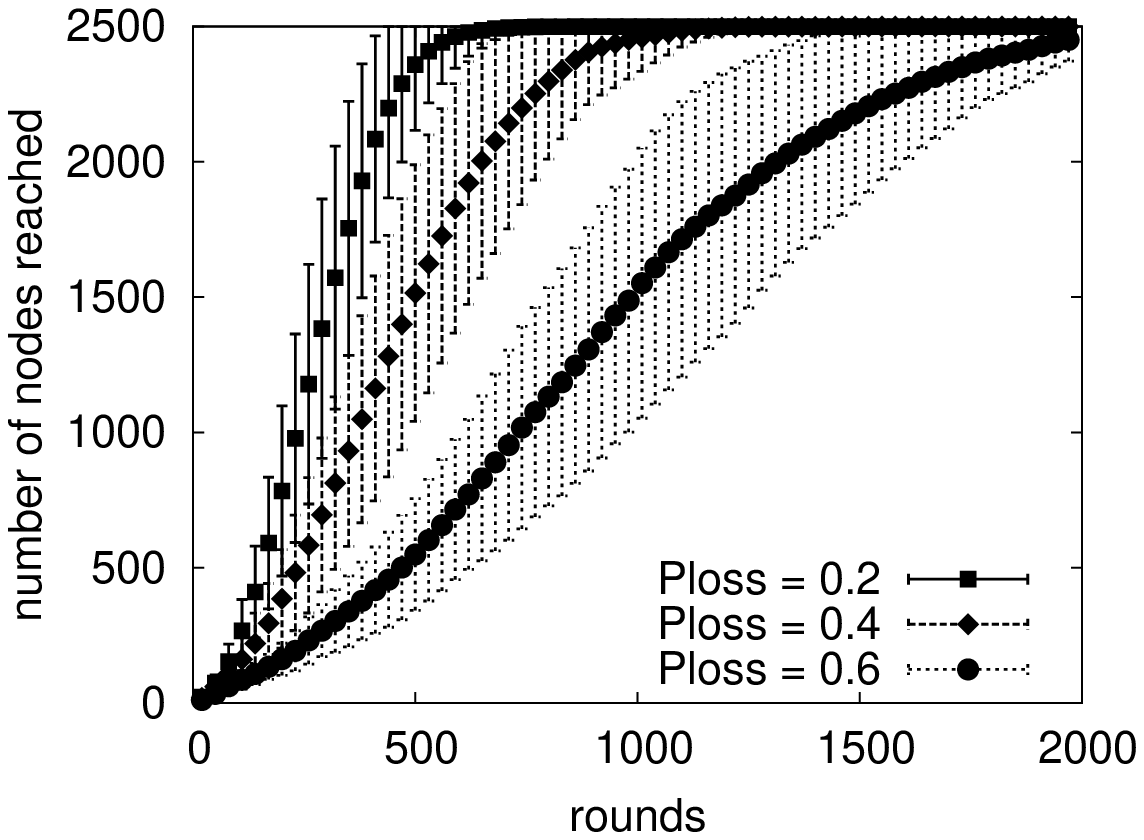} 
        \end{minipage}
        \begin{minipage}[t]{0.49\linewidth}
        \centering\hspace*{-0.6cm}
        \includegraphics[scale=0.5]{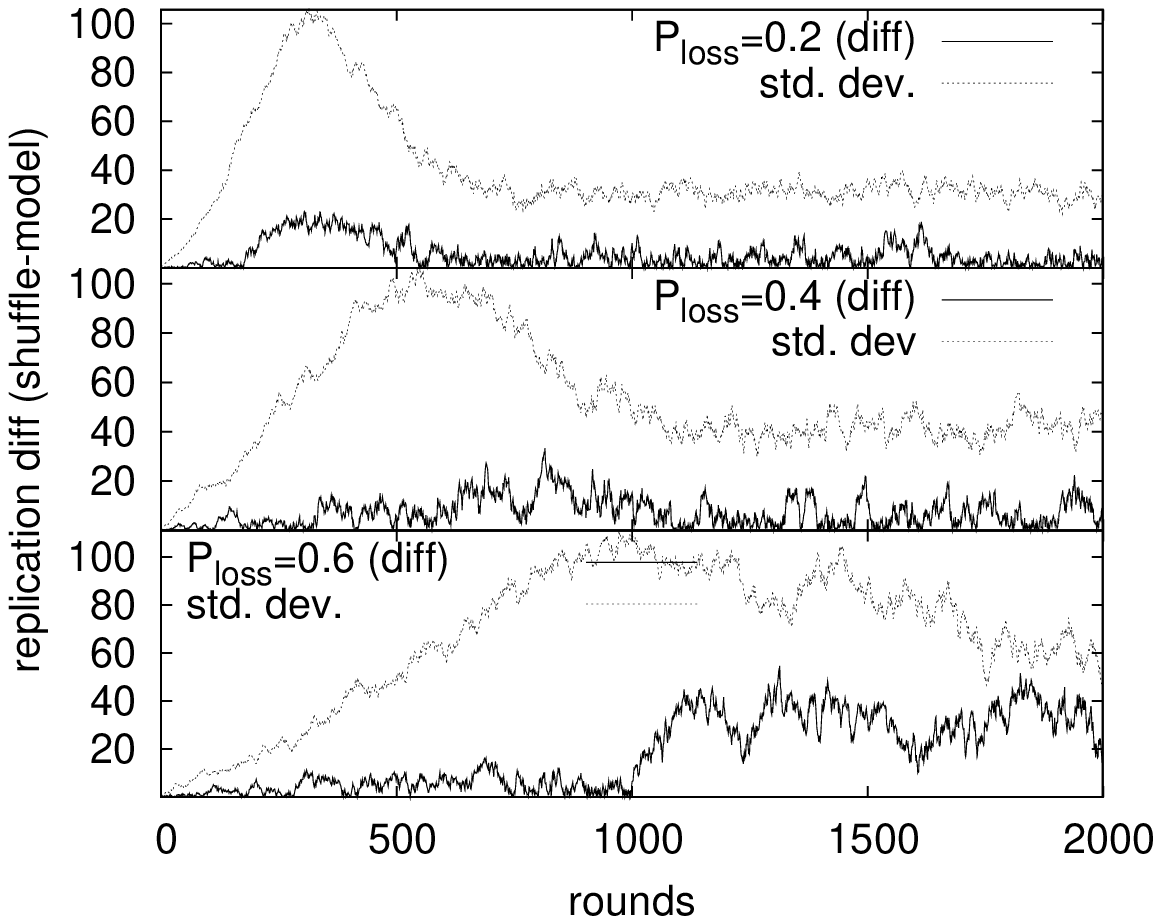}
        \end{minipage}
        \begin{minipage}[t]{0.49\linewidth}
        \centering\hspace*{-0.5cm}
        \includegraphics[scale=0.5]{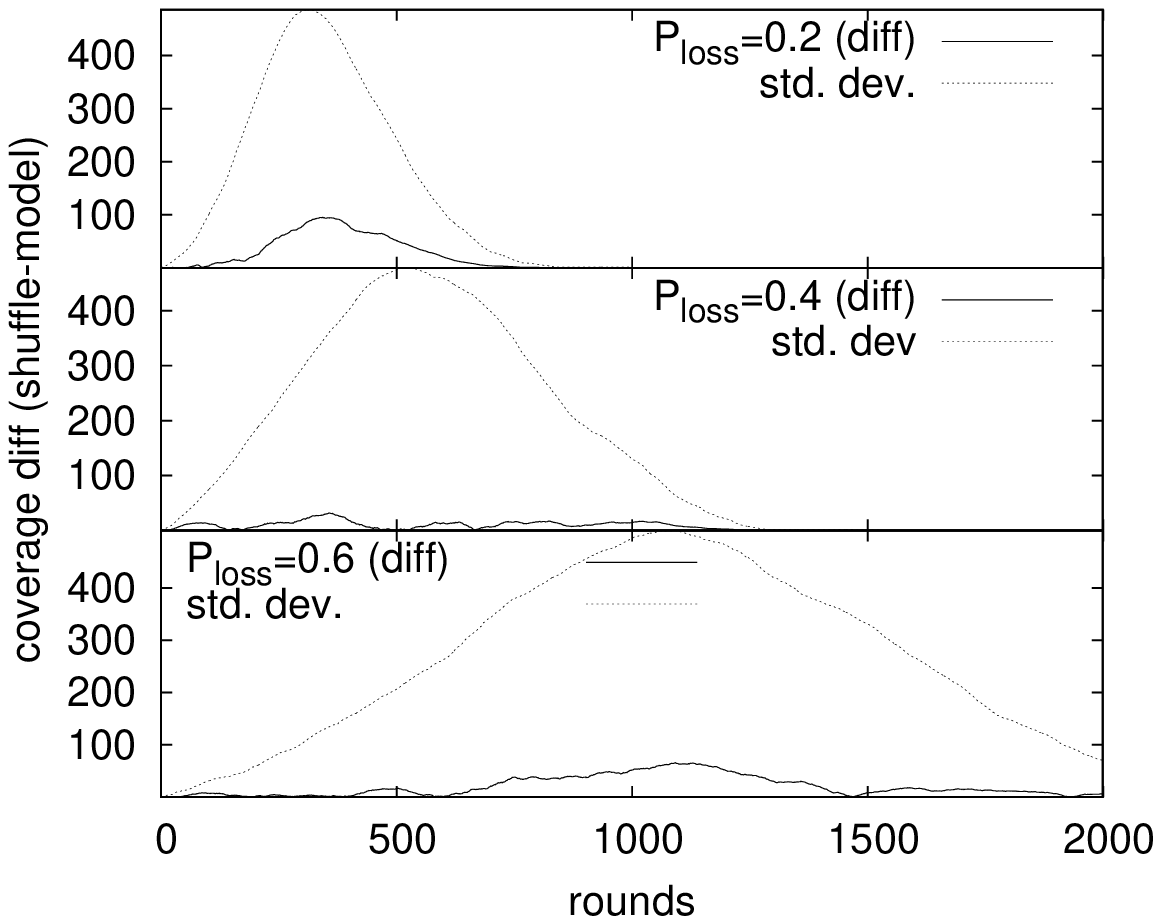}
        \end{minipage}
        \caption{\textbf{Grid, range 1:} Top: Replication (left) and coverage (right) of item 
        $\theitem$ for the shuffle. Bottom: difference between the shuffle and the model 
        for replication (left) and coverage (right).}
        \label{fig:propertiesRange1}
\end{figure} 

\paragraph*{Simulations with the model} For the simulations with the model, $\items, \cache$ and $\buffer$ are system 
parameters set to 500, 100 and 50, respectively. $\proboch$ is determined experimentally on the network, and 
$\probdn$  is calculated using equation \eqref{eq:probdn}. Instead of maintaining a cache, each node in the 
network only maintains a variable that represents whether it holds item $\theitem$ or not (state 1 or 0, 
respectively). Nodes update their state in pairs according to the transition probabilities 
introduced before, see Fig.~\ref{fig:allsc}. This mimics an actual exchange of items between a 
pair of nodes according to Shuffle. While in the protocol this results in both 
nodes updating the contents of their caches, in a simulation using the analytical model 
updating the state of a node refers to updating only one variable: whether the node is in 
possession of item $\theitem$ or not. 
At round 0 we set the state of a random node to 1 (while all the others have state 0) 
and track the state of the nodes for the remainder of the simulation. 

\begin{figure}[!t]
        \begin{minipage}[t]{0.49\linewidth}
        \centering\hspace*{-0.6cm}
        \includegraphics[scale=0.5]{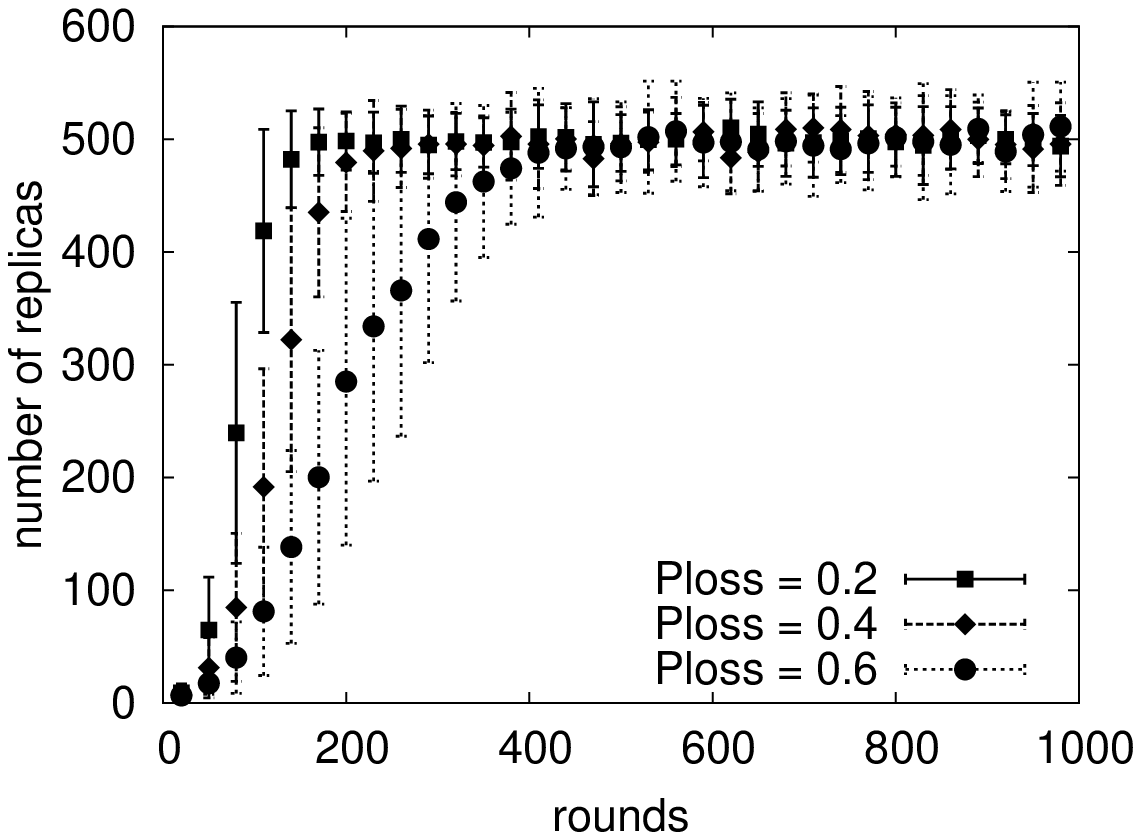}
        \end{minipage}
        \begin{minipage}[t]{0.49\linewidth}
        \centering\hspace*{-0.6cm}        
        \includegraphics[scale=0.5]{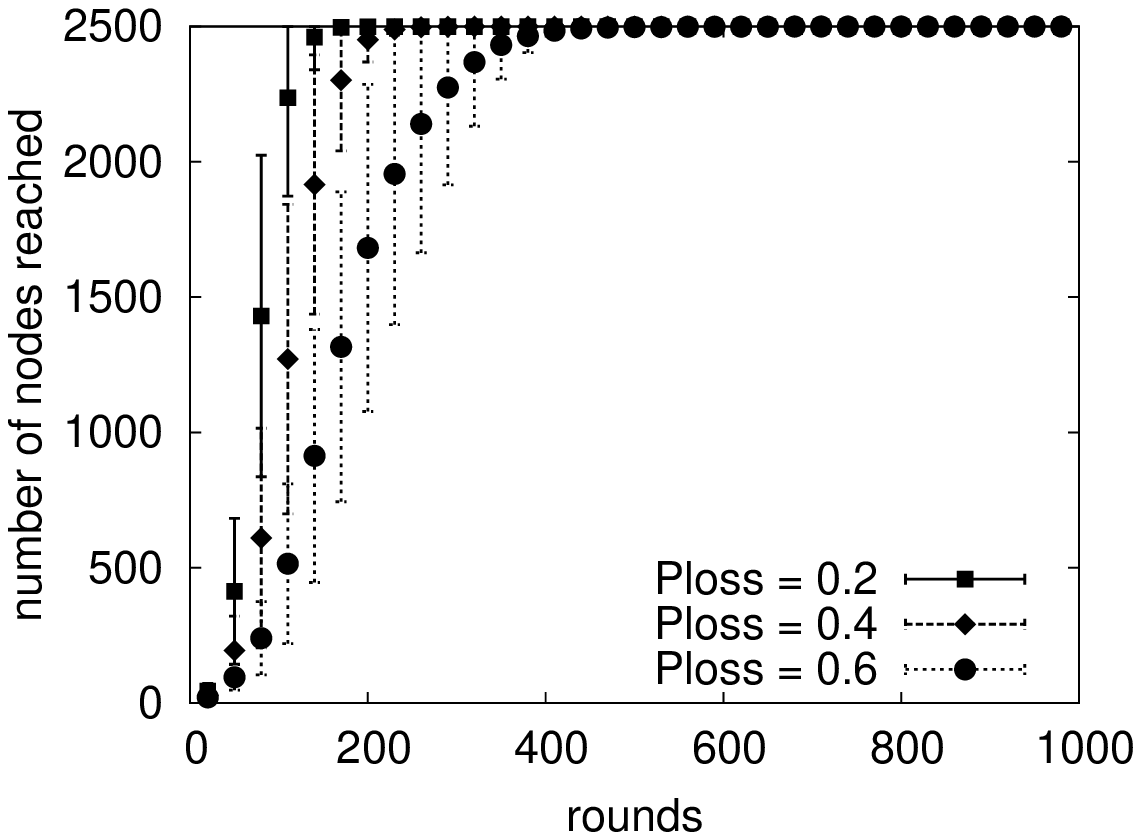}
        \end{minipage}
        \begin{minipage}[t]{0.49\linewidth}
        \centering\hspace*{-0.6cm}
        \includegraphics[scale=0.5]{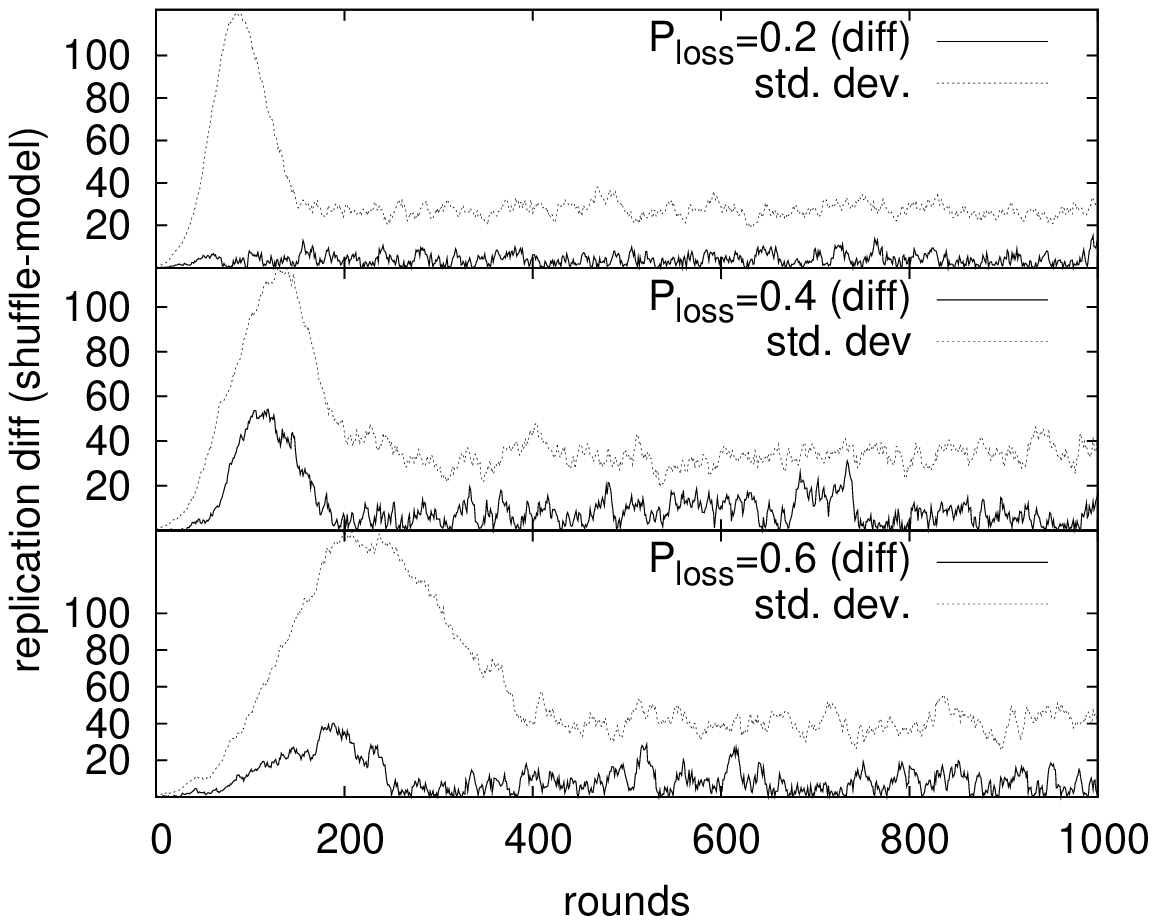}
        \end{minipage}
        \begin{minipage}[t]{0.49\linewidth}
        \centering\hspace*{-0.4cm}        
        \includegraphics[scale=0.5]{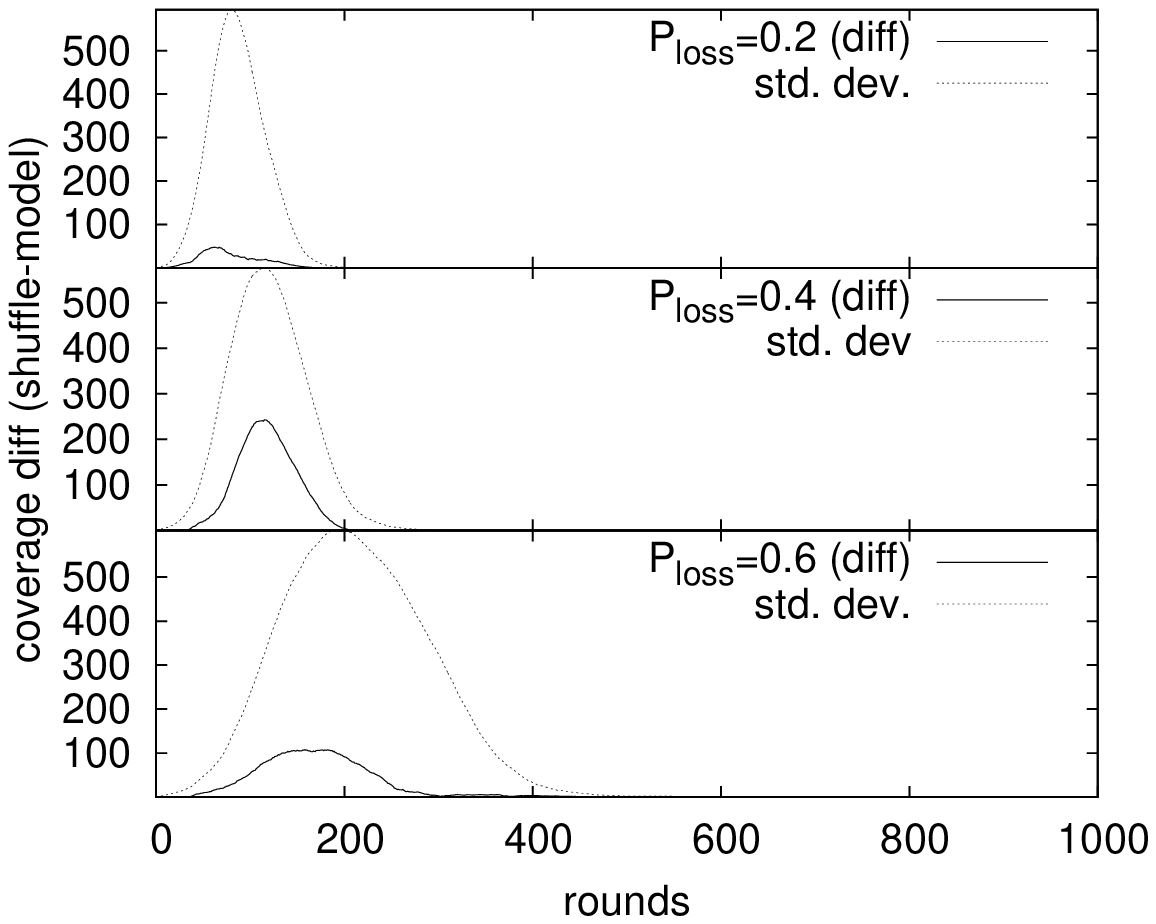}
        \end{minipage}
        \caption{\textbf{Topology with outdegree 4:} Top: Replication (left) and coverage (right) of 
        item $\theitem$ for the shuffle. Bottom: difference between the shuffle and the model for 
        replication (left) and coverage (right).}
        \label{fig:propertiesRG4}
         \vspace{-0.5cm}
\end{figure} 


Figs~\ref{fig:propertiesRange1} and~\ref{fig:propertiesRG4} show the behavior of Shuffle 
(top row) and how it compares to the analytical model (bottom row) in terms of replication (left) and 
coverage (right) of $\theitem$, for different values of $\probls$. For each value of $\probls$, 100 
simulation runs (for both Shuffle and its model) were executed. Due to message loss, it 
is possible for item $\theitem$ to disappear after being introduced into the 
network (usually in the first few rounds). As $\probls$ increases, this situation is more likely. In the 
graphs, we only take into account the successful runs, i.e., where the item did not disappear, but spread. 

The top rows of Figs~\ref{fig:propertiesRange1} and~\ref{fig:propertiesRG4} show the average and standard 
deviation of the successful runs with Shuffle. We compare this data with the average of the 
successful runs of the model, and present the difference (in absolute value) in the bottom rows of 
Figs~\ref{fig:propertiesRange1} and~\ref{fig:propertiesRG4}. We include the standard deviation for the 
shuffle in the bottom row for comparison. It clearly shows that the difference between 
the results obtained from the shuffle and the model fall well within the standard deviation of the 
shuffle results, confirming the ability of the model in predicting the average behavior of the protocol.

In all successful runs (despite message loss), 
the network converges to a situation in which 
there are roughly 500 copies of $\theitem$, indicating that after repeated execution of the protocol 
$\theitem$ receives a fair share of the storage space in the network; $2500 \cdot 100$ cache slots 
divided between 500 items. Also, as expected, due to random gossiping item, $\theitem$ is eventually 
seen by all nodes in the network, when coverage reaches $100\%$.

\section{Broadening the Scope: a Peer-Sampling Protocol}
\label{sec:cyclon}
In the same way that we have developed models for information dissemination algorithms, 
we can apply our methodology to other types of gossip protocols. In this section, we 
describe, in broad terms, how to apply our modelling approach to Cyclon, a gossip-based 
peer-sampling protocol.

\subsection{The Cyclon Protocol}
Peer sampling is a service that nodes in large-scale distributed systems can call to 
obtain a random peer to gossip with. Cyclon implements a peer sampling service 
by constructing and maintaining an unstructured overlay using gossiping 
membership information. 
As opposed to Shuffle, the data items exchanged by Cyclon are references 
to other nodes in the network. The references, or links, that a node stores in 
its cache represent the nodes that it can gossip with. The collection of 
links in the network constitutes an ever-changing overlay over which links are 
exchanged. The aim of Cyclon is to keep the caches of the nodes in the network 
populated with a random selection of links to other nodes. The following table 
summarizes the protocol:

\begin{center}
{\footnotesize
\begin{tabular}{|c|p{8cm}|}
\hline
{\bf method} &  {\bf operation}\\
\hline
\oper{RandomPeer()} & Select $\buffer$ random items and place them 
                      in $\msg$. \\
                    & From $\msg$, select a gossip partner 
                      uniformly at random.\\
\hline
\oper{PrepareMsg()} &  If the node initiates the exchange, 
                       replace the link to the gossip partner 
                       with a link to itself in $\msg$. Return $\msg$.\\
\hline
\oper{Update($\msg,{\msg}_p$)} & Discard links to the node itself and items 
                                 that are already in the cache from ${\msg}_p$. \\
                               & Include remaining items in cache by 
                                 1) using empty slots and 2) replacing 
                                 entries among the ones in $\msg$.  \\
\hline
\end{tabular}
} 
\end{center}

Note that the steps taken within the routines of Cyclon depend on the role that 
a node has in the gossip exchange. The node that initiates the exchange always 
includes a link to itself in the message, while the contacted node does not. This 
is reflected in the \oper{PrepareMsg()} routine, which is slightly different 
for the initiator.

\subsection{Modelling Cyclon: States and Transitions}
According to Cyclon, every link $\theitem$ has a publisher $D$, and whenever $D$ gossips, 
it sends the link $\theitem$ to its partner. Thus, when modelling the 
dissemination of link $\theitem$ throughout the network, we have to take into account 
whether $D$ (the source of link $\theitem$) is involved in the exchange. Therefore, we 
define the state of a node as a two-bit string, where the first bit indicates 
whether the node is $D$ and the second bit indicates whether link $\theitem$ is 
in the node's cache (as done in the previous models).

As nodes update their states in pairs, and with each node having a state represented 
by two bits, the state of a pair of nodes engaged in a Cyclon exchange consists of 
a four-bit string. While this opens up to the possibility for 16 states, only 8 
states are valid. Cyclon precludes a node from storing its own link, eliminating 
seven states (where one of the nodes has state $11$). In addition, the state 
$1010$, where $D$ contacts itself, is also invalid.

\begin{figure}[!t]
 \centering
\begin{minipage}[t]{0.45\linewidth}
  \centering
     \includegraphics[scale=0.9]{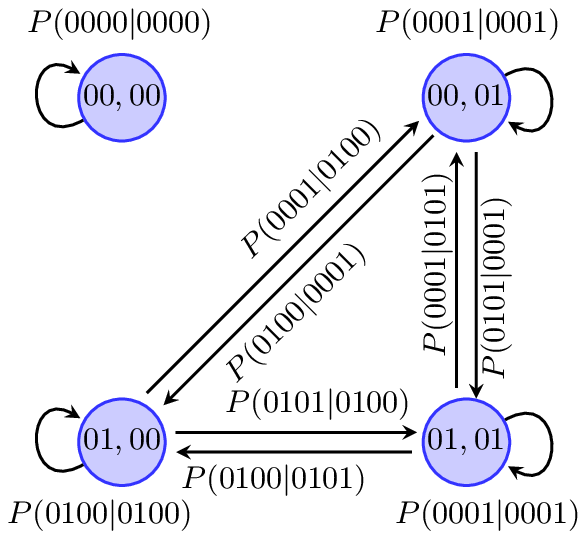}
 \caption{State transitions when node A and B are not producers of item $\theitem$.}
 \label{fig:statesWithoutNodeD}
\end{minipage}
 \hspace{0.4cm}
\begin{minipage}[t]{0.45\linewidth}
  \centering
     \includegraphics[scale=0.9]{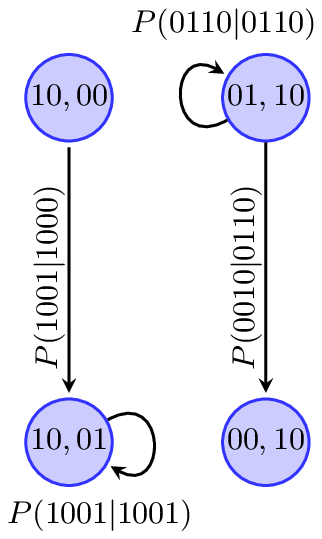}
  \caption{State transitions when either node $A$ or $B$ is the producer of item $\theitem$.}
  \label{fig:statesWithNodeD}
\end{minipage}
\vspace{-0.3cm}
\end{figure}

Fig.~\ref{fig:statesWithoutNodeD} shows the transitions between states where 
$D$ is not involved. For simplicity, we assume that there are no link failures, 
resulting in a transition diagram similar to what we would encounter for Shuffle. 
Note that the four states depicted have the form 0X0X, indicating that 
neither the initiator nor the gossip partner is $D$.

When $D$ is involved in the exchange, there are only four possible states, 
as shown in Fig.~\ref{fig:statesWithNodeD}. Since $D$ as the initiator always 
includes a link to itself in the set of items sent, the outcome will always be 
that the other node has a link $\theitem$. On the other hand, when $D$ is the 
contacted node (see bottom of figure), the initiator might or might not drop 
link $\theitem$ from its cache as a result. In all cases, $D$ always keeps the 
state $10$, as it can never keep a link to itself in its cache. 
%

Describing the transitions in terms of the building blocks $\probsn$ and $\probdn$ can 
be done in a similar way as with Shuffle (when $D$ is not involved in the 
exchange). The variations in the routines, specifically {\sf\small PrepareMsg()}, 
due to the role taken by the node in the exchange, have to be given special consideration, 
as they will cause $\probsn$ to be slightly different for the initiator and the 
contacted node. When node $D$ is involved, Cyclon is very clear about the outcome. 
The transitions can be easily calculated. If node $D$ is the initiator, there is 
only one possible outcome (probability = 1). If node $D$ is the contacted node, 
the transitions depend solely on $\probdn$.

\subsection{Other Protocols}

As mentioned before, in the original version of Newscast, 
data items are timestamped according 
to their creation time. A node allocates the space in its cache for 
the received items by discarding corresponding number of oldest items. 
If an item is received that is also in the node's cache, the recent 
version of the item is preferred. 
Like for Cyclon, a state of the transition diagram of Newscast is then 
a pair of tuples $(a,b)$, where $a,b \in \{0,(1,i)\}$. The first bit 
indicates if the data item is present in the gossiping node's cache. The 
integer $i$ is a timestamp of the corresponding item. The expression for 
probability $\probsn$ remains the same, but $\probdn$ is then a function 
of the corresponding timestamp. It can be either derived through rigorous 
analysis, or can be measured experimentally similar to the approach presented 
in Sec.~\ref{sec:lossy}.

\section{Conclusions}
\label{sec:conclusion}

Traditional analysis methods for computer systems such as model checking with its exhaustive state space 
search, fail to cope  with large networks. Although quite useful for studying certain 
behavior of small-sized networks, these methods do not scale well for \gp{} 
protocols. On the other hand, standard methods for modelling of epidemics 
scale well, but abstract away the details of a protocol, showing only simple emergent 
behavior of the system. A challenge is to develop analytical models that 
capture (part of) the behavior of a system, and then subsequently optimize 
design parameters, at the right level of abstraction.

In this paper, we presented an analytical framework for a class of gossip-based information 
dissemination protocols. 
Modelling a gossip protocol at the level of local pairwise interactions, as demonstrated 
in this paper, is a scalable approach to analyze such protocols. On the one hand, 
this analytical model includes the mechanics of communication between nodes on 
the level of the protocol details. On the other hand, such a model allows for 
studying the impact of system parameters on the performance of the protocol, and 
can be used to optimally design and fine-tune it.

Furthermore, we have presented the analysis of a gossip protocol 
in the presence of transient communication failures. For our framework, we introduced a 
hybrid method to compute $\probdn$ that combines both rigorous modelling of the protocol and statistical 
data sampling from large-scale Monte Carlo simulations for different network topologies. To be more 
precise,  we derive analytical expressions for components of our framework that are invariant with 
respect to the scenarios we want to model. Having decomposed our model into its basic components, 
we identify the ones that are affected by specific scenario configurations (in this case, message 
loss and topology combinations). Finding analytical expressions for these components (in this work, 
only $\proboch$) would require modelling each specific scenario configuration, reducing the applicability 
of the framework to those very specific configurations. Instead, we isolate the framework from the 
scenario-specific component and use statistical data sampling to obtain a value for it.

We opt for this combined approach since it allows us to build a framework that captures the behavior 
of the \gp{} protocol without requiring us to incorporate scenario-specific elements into 
the model. The challenge in this approach lies in being able to decompose the model into its 
minimal components, in such a way that the ones which are dependent on particular scenarios 
(for which expressions that encompass all scenarios cannot be derived) can be isolated and 
computed separately. In effect, we strive for a golden mean between high-level models such as 
for epidemics showing only the emergent behavior and the low-level models of the protocol that 
depend on particular implementation settings. 

With respect to gossip-based dissemination, our study revealed that for networks with 
lossy communication channels, the assumption of a uniform distribution of data is valid only if every node 
can gossip with any other node in the network.
\revision{In future, we plan to study the impact of different gossiping frequency on the performance results.}

\bibliographystyle{abbrv}
\bibliography{icdcs}
\appendix
\section{Pull-only and push-only variations}
\label{sec:onedir}
Our framework is not limited to push-pull gossip protocols, but allows for modelling of push-only as well 
pull-only gossip protocols. To illustrate this, we use two variations of the simplified Newscast, and 
briefly demonstrate the calculation of transition probabilities for push and pull versions of the protocol.

\subsection*{Push-only protocol}
Recall that in push-based gossip protocols, only the initiator node $A$ sends its message to $B$, but does not 
update its own cache. Therefore, the transitions where $A$ changes its state are not possible anymore, 
and the probability of such transitions is 0. The transition probability $P(00|00) = 1$ since items do not 
appear from nowhere.

\begin{scenario}[$0,1$] Before gossip, only $B$ has $\theitem$ in its cache.
\begin{compactdesc}
 \item[$a_2 b_2=01$:] node $B$ does not drop $\theitem$ from its cache; 
  \(P(01|01)= \probkn \).
 \item[$a_2 b_2=00$:] node $B$ drops $\theitem$ during the update of its cache; 
 \(P(00|01)= \probdn\).
\end{compactdesc}
\end{scenario}
\begin{scenario}[$1,0$] Before gossip, $\theitem$ is in the cache of $A$.
\begin{compactdesc}
 \item[$a_2 b_2=10$:] node $A$ either does not send $\theitem$, or sends it, but $B$ drops $\theitem$, \ie,\ the 
 probability is \(P(10|10)= \probnotsn + \probsn \cdot \probdn \).
 \item[$a_2 b_2=11$:] node $A$ sends the item $\theitem$, and node $B$ keeps it: \(P(11|10)= \probsn \cdot \probkn\).
\end{compactdesc}
\end{scenario}
\begin{scenario}[$1,1$] Before gossip, $\theitem$ is in the caches of both nodes.
\begin{compactdesc}
 \item[$a_2 b_2=10$:] node $B$ drops $\theitem$ from its cache during the update; \( P(10|11) = \probdn \).
 \item[$a_2 b_2=11$:] node $B$ keeps the item $\theitem$ after the update; \ie, \(P(11|11)= \probkn \).
\end{compactdesc}
\end{scenario}

\subsection*{Pull-only protocol}

In contrast to the push version, in pull-based gossip protocols the initiator node 
only requests data from its peer. In terms of state transitions, only initiator $A$ can 
update its state, but not the contacted node $B$. Thus, the probability of 
transitions in which the state of $B$ before gossip differs from its state after 
the gossip is 0. The state $(0,0)$ again has a self-transition with probability 1: 
$P(00|00) = 1$.

\begin{scenario}[$0,1$] Before gossip, only $B$ has $\theitem$ in its cache.
\begin{compactdesc}
 \item[$a_2 b_2=01$:] node $A$ does not get $\theitem$ from $B$, or $B$ sends the item, 
 but $A$ drops it; 
  \(P(01|01)= \probnotsn + \probsn \cdot \probdn \).
 \item[$a_2 b_2=11$:] node $B$ sends $\theitem$ and node $A$ keeps it; 
 \(P(11|01)= \probsn \cdot \probkn \).
\end{compactdesc}
\end{scenario}
\begin{scenario}[$1,0$] Before gossip, $\theitem$ is in the cache of $A$.
\begin{compactdesc}
 \item[$a_2 b_2=10$:] node $A$ does not drop $\theitem$ after its cache update, \ie,\ the 
 probability is \(P(10|10)= \probkn \).
 \item[$a_2 b_2=00$:] node $A$ does not keep the item for its updated cache: \(P(00|10)= \probdn\).
\end{compactdesc}
\end{scenario}
\begin{scenario}[$1,1$] Before gossip, $\theitem$ is in the caches of both nodes.
\begin{compactdesc}
 \item[$a_2 b_2=01$:] node $A$ drops $\theitem$ from its cache during the update; \( P(01|11) = \probdn \).
 \item[$a_2 b_2=11$:] node $A$ keeps the item $\theitem$ after the update with 
 the probability \(P(11|11)= \probkn \).
\end{compactdesc}
\end{scenario}

\end{document}